\documentclass[12pt]{article} 
\usepackage{array}
\usepackage{makecell}
\usepackage{booktabs}
\usepackage{amssymb}
\usepackage{bm}
\usepackage{mathrsfs} 
\usepackage{natbib}
\usepackage{algorithm}
\usepackage{algpseudocode}
\usepackage{hyperref}
\usepackage{amsthm}
\usepackage{mathtools}
\usepackage{epstopdf,xcolor}
\usepackage{bbm}
\usepackage[colorinlistoftodos,prependcaption]{todonotes}
\usepackage{dsfont}
\usepackage{pifont}
\usepackage{subcaption}
\usepackage{tikz, subcaption}
\usetikzlibrary{bayesnet, shapes, arrows, positioning}
\usepackage{fullpage}
\usepackage{enumitem}
\usepackage{graphicx}
\usepackage[page]{appendix}
\usepackage{comment}
\usepackage{multirow}
\usepackage{authblk}
\usepackage{listings}
\usepackage{booktabs}
\usepackage{siunitx}

\newtheorem{theorem}{Theorem}

\newtheorem{proposition}{Proposition}

\newcommand{\E}{\mathbb{E}}
\newcommand{\R}{\mathbb{R}}

\newcommand{\tr}{\mathrm{tr}}
\newcommand{\cov}{\mathrm{cov}}
\newcommand{\etr}{\mathrm{etr}}

\title{\bf \Large Spatially orthogonal factor models\\for spatial transcriptomics and remote sensing data}

\author[1]{Dan Cunha}
\author[2]{Lukas M. Weber}
\author[3]{Mark Friedl}
\author[1]{Luis Carvalho}
\affil[1]{Department of Mathematics and Statistics, Boston University}
\affil[2]{Department of Biostatistics, Boston University School of Public Health}
\affil[3]{Department of Earth and Environment, Boston University}
\date{}

\begin{document}
\maketitle

\begin{abstract}
Principal component and factor analyses are often applied to spatial data towards inference on latent modes of spatial variation. These analyses are widespread across domains including spatial transcriptomics and environmental/climate sciences, where the modes of spatial variation are represented by corresponding factors of gene expression or remotely sensed time series measurements. Many methods have been proposed for incorporating spatial information into a probabilistic PCA framework; however, there are three main drawbacks to currently available approaches. First, the loadings matrices in the probabilistic PCA model specification are not orthogonal, and subsequent orthogonalization of those loadings corrupts the original prior spatial information. Furthermore, currently proposed methods assume stationarity in their spatial prior and either fix the prior or use a rule-of-thumb estimator. Finally, dimensionality reduction methods that incorporate spatial information typically do not achieve linear-time computational complexity with respect to the number of spatial locations. To resolve these problems, we first parameterize the model directly with orthogonal loadings. For their prior distribution, we derive the sampling distribution of a singular value decomposition transformation with $k$ unique singular values and $m-k$ repeated singular values. We then show under this model that the maximum a posteriori estimator for the orthogonal loadings is the eigendecomposition of $S + \frac{1}{n}\Sigma$, where $S$ is the empirical covariance matrix and $\Sigma$ is the prior spatial covariance. We develop a minorization-maximization-within-EM algorithm that is linear in computational complexity with respect to the number of spatial locations. We further extend our MM-EM algorithm to handle held-out locations and develop a validation strategy for optimizing the nonstationary prior covariance. Our methodology is used to infer the spatial distribution of direction-specific length scales in a human brain spatial transcriptomics case study, as well as a continental-scale phenology case study in sub-Saharan Africa.
\end{abstract}

\newpage
\section{Introduction}
\label{sec:intro}
In the analysis of spatially indexed data including biological data from spatial transcriptomics technologies and remote sensing data from environmental/climate sciences, there is a demand for answering statistical questions of the following kind.  Which combinations of gene expression explain the major modes of spatial variation across biological tissue? Or, what environmental temporal patterns represent important modes of spatial variation geographically? These questions have an important commonality; the corresponding data are spatially indexed, and we expect, \textit{a priori}, the modes of variation to be spatially distributed.  

There are a number of challenges incorporating spatial information into dimensionality reduction models: 
\begin{itemize}
    \item The loadings matrices in the probabilistic principal component analysis (PCA) model specification (introduced below) are not orthogonal, and subsequent orthogonalization of those loadings corrupts the original prior spatial information.
    \item Currently proposed methods assume stationarity in their spatial prior and either fix the prior or use a rule-of-thumb estimator.
    \item Dimensionality reduction methods that incorporate spatial information typically do not achieve linear-time computational complexity with respect to the number of spatial locations.
    \item It is unclear how the incorporation of spatial information changes the resulting maximum likelihood or maximum a posteriori inferences.
\end{itemize}

The purpose of this work is to build upon probabilistic PCA to address the above challenges. 
\subsection{Probabilistic principal component analysis}
Principal component analysis (PCA) is a tool found ubiquitously throughout scientific application \citep{pearson1901liii,hotelling1933analysis}.  % ---Introduce probabilistic PCA model
A probabilistic characterization of PCA follows \cite{tipping1999probabilistic}.  Suppose $Y$ is an $n\times m$ matrix, where the rows index observations or features, and the columns index spatial locations. In environmental applications, rows typically correspond to time points (referred to as `observations'), while in spatial transcriptomics rows typically correspond to genes (referred to as `features'). In this work, in both applied settings, we will use the term `observations' for the rows of $Y$ and spatial locations for the columns. (In addition, we note that in the spatial transcriptomics literature, `observation' may refer to a spatial location together with its vector of measurements across all genes / features; we do not use this terminology here.) We apply the model after removing application-specific mean structure. In the remote sensing analysis, each spatial location, corresponding to a column of $Y$, is mean-centered across temporal observations. In the spatial transcriptomics analysis, each gene, corresponding to a row of $Y$, is mean-centered across spatial locations (after initial preprocessing including normalization and transformation). For notational clarity, we assume $Y$ is already mean centered. Let the $i$th row of $Y$ be denoted as $\bm{y}_i$. Under the probabilistic PCA model with $k$ components, $\bm{y}_i$ has the following independent conditional likelihood distribution for each $i$th observation,
\[
\bm{y}_i|W,\sigma^2,\bm{z}_i \sim N(W\bm{z}_i, \sigma^2I_m)
\]
where $W \in \R^{m\times k}$ is the loadings matrix and $\bm{z}_i$ represents the $k$ latent factors for the $i$th observation,
\[
\bm{z}_i \sim N(0,I_k)
\]
independently for each $i = 1,\dots,n$. When $\bm{z}_i$ is marginalized out, the marginal likelihood of $y_i$ follows,
\[
\bm{y}_i |W,\sigma^2 \sim N(0, WW' + \sigma^2 I_m)
\]
\cite{tipping1999probabilistic} showed that the maximum likelihood estimator for $W$, in terms of its singular value decomposition $ULV'$, is such that $\hat{U}$ are the top-$k$ eigenvectors of the empirical covariance matrix $S \coloneqq Y'Y/n$, and $\hat{L}$ is related to the eigenvalues of $S$. But what if we have additional spatial information about the proximity of the $m$ locations? For example, perhaps we have a spatial covariance matrix $\Sigma$ with proximal information about locations across space. How should that information be included in the model, and how does that incorporation change our inference?

\subsection{Incorporating prior spatial information}
\subsubsection{Prior work}
There have been a number of contributions towards incorporating spatial information in PCA or factor analyses in the biological and environmental sciences, and we only review a small number of them here. For a more comprehensive review in the environmental and climate sciences, the reader is referred to \cite{hannachi2007empirical}. \cite{green1988transformation} study the effect of preprocessing data, using spatial transformations, on the ordering of principal components and their spatial quality.  \cite{switzer1984tecnical} develop min/max autocorrelation factors to order components according to their spatial autocorrelation. In the biological setting on the analysis of spatial transcriptomics data, \cite{shang2022spatially} treat the spatial dimension as factors and assume a stationary spatial covariance in their prior distribution. \cite{townes2023nonnegative} incorporate spatial information in the factor prior within a probabilistic nonnegative matrix factorization. \citet{guo2026jade,PASTE_zeira2022alignment,sedr_xu2024unsupervised,BAYESSPACE_zhao2021spatial,stlearn_pham2020,spagcn_hu2021,deepst_xu2022,dong2022deciphering} have provided solutions to tasks such as spatial clustering, trajectory inference, spatially variable genes, and tissue sample alignment, each of which use spatial latent embeddings within their solutions. 

\subsubsection{Considerations for incorporating prior spatial information}
There are two high-level points to deliberate: should the loadings or factors act as the spatial dimension and, should spatial dependence be incorporated into the mean or the errors of the model?  Regarding the first question, we follow decades of environmental science research placing the spatial dimension as loadings and the time/gene dimension as factors. As such, the burden is to defend this decision within the biological setting in the context of spatial transcriptomics data. Our reasoning is as follows. We wish to infer orthogonal spatial patterns (that is, spatial patterns that are maximally different in the sense that their inner product is $0$) associated with gene expression combinations, which may not themselves be orthogonal, as opposed to inferring orthogonal gene expression combinations that may share similar spatial patterns.
%We wish to infer Euclidean-orthogonal spatial patterns associated with gene combinations, which may not themselves be orthogonal, as opposed to inferring orthogonal gene combinations that may share similar spatial regions. Here, orthogonality is defined across spatial locations and reduces redundancy among the inferred spatial patterns; it does not imply that the corresponding biological processes are independent or nonoverlapping. In spatial transcriptomics data, we expect some dominant gene expression programs to be organized by tissue anatomy, cell-type composition, immune-associated expression response, or local cellular neighborhoods.
In spatial transcriptomics data, we expect that some dominant gene expression programs may overlap in their constituent genes and exhibit spatial organization associated with tissue anatomy, cell-type composition, immune-related processes, or local cellular neighborhoods. %while residual variation may reflect technical effects including spot-level sampling and capture efficiency, and cell-mixture effects.

Regarding the second question, our options are to place spatial dependence within the error distribution or within the loadings distribution. The former has been extensively studied in the environmental science literature \citep{hannachi2007empirical}. In our case study, however, we expect phenological modes of variation to manifest spatially, more so than the errors from the conditional mean function. Furthermore, the GIMMS dataset we are using is relatively robust to errors due to cloud cover compared to other satellite imagery, as it is a 15-day maximum value composite of daily imagery \citep{fensholt2012evaluation}.

% ---Introduce a spatial prior on W.
% ---Motivate that since researchers want orthogonal loadings, and spatial smoothness, so we tackle this problem directly, avoiding the problem that it is unclear how spatial properties on W change as a result of orthogonalization and rotation.
As such, we proceed by placing spatial dependence on the loadings.  However, there is another issue to address. If we place a prior spatial dependence on the columns of the (weighted and rotated) loadings matrix $W$, then it is unclear how those prior assumptions are transformed if our final inferences are on $U$, the orthogonal basis of its column space after singular value decomposition.  To address this issue, we model $Y$ using the orthogonal matrix $U$ directly, such that
\begin{equation}
\label{eq:marg_lik}
    \bm{y}_i |U,L,\sigma^2 \sim N(0, UL^2U' + \sigma^2 I_m)    
\end{equation}
%Under these assumptions, we derive a the sampling distribution for $U,L,\sigma^2$.

\subsection{Contributions}
Our contributions are as follows:
\begin{itemize}
    \item We derive the sampling distribution for the singular value decomposition when the first $k$ singular values are unique and the remaining $m-k$ are all equal (Proposition \ref{prop:prior} in Section \ref{sec:sofm}).
    \item We place this distribution on the orthogonal loadings $U$ and their corresponding singular values $L$, and show that the corresponding maximum \textit{a posteriori} estimate of $U$ consists of the eigenvectors of $S + \frac{1}{n}\Sigma$, where $S$ is the empirical covariance matrix and $\Sigma$ is the prior spatial covariance matrix (Theorem \ref{thm:MAP_U} in Section \ref{sec:sofm}).
    \item As the eigendecomposition of $S + \frac{1}{n}\Sigma$ has a cubic computational complexity with respect to the number of spatial locations $m$, in Section \ref{sec:estimation} we derive a minorization-maximization-within-EM (MM-EM) algorithm for estimating the orthogonal loadings $U$, achieving a linear-time computational complexity with respect to the number of spatial locations.
    \item In Section \ref{sec:validation}, we extend our methodology to handle held-out spatial locations and subsequently maximize the held-out predictive distribution with respect to the prior spatial covariance on the orthogonal loadings $U$ under a spatial smoothness assumption.
    \item In Section \ref{sec:DLPFC}, we study spatial transcriptomics data from a postmortem human brain tissue section from the dorsolateral prefrontal cortex (DLPFC) region, and identify spatial factors associated with laminar and non-laminar gene expression patterns that are consistent with findings in the existing literature.
    \item In Section \ref{sec:Africa}, we apply our model and methodology to study phenological modes of spatial variation in sub-Saharan Africa.
\end{itemize}

\section{Spatially orthogonal factor models}
\label{sec:sofm}
\subsection{Prior genesis of spatially orthogonal factor loadings}
% sub: Prior genesis of spatially orthogonal factor loadings
% ---But to model U,L,sigma directly, we need careful jacobian accounting
% ---Proposition: prior on U,L,sigma is such and such
% ---Note taking account of repeated singular values 
% ---Note normalizing constant intractable which motivated our validation approach for prior estimation
% ---This is a matrix fisher bingham
% ---We derive both cases when dof is less than m in the singular case, and greater than or equal to m
We proceed with deriving the prior sampling distribution for $U,L,\sigma^2$, such that the spatial dependence on the loadings follows a covariance matrix $\Sigma$. We will specify the form of $\Sigma$ in Subsection \ref{sub:spatial_cov} as well as a methodology for estimating $\Sigma$ in Section \ref{sec:validation}. Our results build off the seminal results due to \cite{james1964distributions}, for the sampling distribution of an eigendecomposition of a Wishart distributed random matrix. However, to align with PCA assumptions, we restrict the eigendecomposition such that the first $k$ eigenvalues are unique and the remaining are repeated. We then take an inverse transformation to achieve conjugacy with the likelihood and reparameterize the eigenvalues to align with PCA parameterization, such that $L_{jj}^2 + \sigma^2$ are the first unique $k$ eigenvalues and $\sigma^2$ are the remaining $m-k$ eigenvalues.
\begin{proposition}(Prior genesis)
\label{prop:prior}
Suppose the prior density $p(Y^{(0)}) \propto \etr(-\frac{1}{2} Y^{(0)'} \Sigma Y^{(0)})$, is limited to the sample space where $Y^{(0)}$ has $k$ unique singular values as described above. Then the three Jacobian transformations described above yield the distribution,
\begin{align*}
    p(U,L, \sigma^2|\Sigma) &\propto \exp{\left(-\frac{1}{2}\tr(U \left((L^2 + \sigma^2 I_k)^{-1} - \sigma^{-2}I_k\right) U' \Sigma + \frac{1}{\sigma^2}\Sigma)\right)}\\
    &\quad \prod_{j>i}^k \left((L_{jj}^2 + \sigma^2)^{-1} - (L_{ii}^2 + \sigma^2)^{-1}\right)
     \left(\prod_{j=1}^{k} ((L_{jj}^2 + \sigma^2)^{-1} - \sigma^{-2})^{(m-k)} \right)\\
     &\quad \left((\sigma^{2*\frac{1}{2}})^{-(\nu-m)(m-k)-\binom{m-k}{2}-3}\right)
     \left(\prod_{j = 1}^k (L_{jj}^2 + \sigma^2)^{-\frac{(\nu-m)+3}{2}}\right)\\
     &\quad \frac{1}{2} \det(L) (dL)(d\sigma^2)(dU)
\end{align*}
Specifically, the conditional prior of $U|L,\sigma^2, \Sigma$ is a matrix-Fisher-Bingham and conjugate to the likelihood. 
\end{proposition}
We use exterior product theory for our proof following \cite{muirhead2009aspects}. As we will see, these transformations yield an interpretable maximum \textit{a posteriori} inference of $U$.

There are a few parameters that control the strength of this prior.  The first is the degrees of freedom parameter $\nu$ which we take to be equal to $m$; the smallest value $\nu$ can take yielding the above distribution. The second is the variance parameter of $\Sigma$, and the third is the length scale. The parameters of $\Sigma$ have a strong influence on the posterior results for $U$, but we cannot optimize these directly as the normalizing constant of this prior is difficult to compute. We develop a validation methodology for estimating $\Sigma$ in Section \ref{sec:validation}.

\subsection{MAP inference of spatially orthogonal factor loadings}
% sub: MAP Inference of spatially orthogonal factor loadings
% ---Motivate ideally that we want to know S + 1/n Sigma, explain other methods do not know the analytic form of their estimator
% ---Motivate that matrix fisher bingham is conjugate to marginal likelihood
% ---Theorem: map of U are the eigenvectors of S + 1/n Sigma
% 	---also note singular wishart does not satisfy this property, thus dof need to be >= m
% ---One key takeaway, outside of interpretability, is that once can infer k_max and do not need to refit model for all k = 1,...,k_max
% ---Regarding L
% ---Note corollary in appendix to Tipping Bishop: the map of Wishart of a specific form is equivalent to Tipping Bishop claim
% ---Note map of L are the eigenvalues of S + 1/n Sigma if you reverse Jacobian back to W, then argmax W, then SVD 
% ---Note that otherwise it is an open problem we are working to understand
Proposition \ref{prop:prior} establishes the ground work for the following theorem. By deriving the sampling distribution of $U,L,\sigma^2$ under the repeated eigenvalue constraint and reparameterizing according to PCA assumptions, conjugacy is established and yields an interpretable MAP inference,
\begin{theorem}(MAP inference)
\label{thm:MAP_U}
    Let the likelihood distribution of the data be as defined in Equation \ref{eq:marg_lik}, and let the prior on $U,L,\sigma^2$ be as defined in Proposition \ref{prop:prior} for $\nu \geq m$. Then the unconditional maximum \textit{a posteriori} (MAP) of $U$,
    \[
    U^\dagger = \underset{U}{\mathrm{argmax}}\ p(U,L,\sigma^2 | Y,\Sigma)
    \]
    are the $k$ eigenvectors of $(S + \frac{1}{n}\Sigma)$ associated with the corresponding largest $k$ eigenvalues.
\end{theorem}
This MAP inference is unconditional on the values of $L,\sigma^2$ in the following sense.  We assume without loss of generality that the diagonal values of $L$ are descending in value and $\sigma^2>0$. Under these assumptions, the MAP for $U$ does not otherwise depend on the values of $L,\sigma^2$.  The MAP values for $L,\sigma^2$ are more difficult to prove and we leave this step for potential future research.

\subsection{Spatial covariance models}
\label{sub:spatial_cov}
In this subsection, we describe the nonstationary covariance function used in our studies. We start by specifying an anisotropic squared exponential stationary function which will be used in the initial stages of our nonstationary estimation procedure,
\[
C_s(s_i, s_j) = \phi\cdot\etr\left(-\frac{1}{2}(\bm{s}_i - \bm{s}_j)'\Sigma^{-1} (\bm{s}_i - \bm{s}_j)\right)
\]
where,
\[
\Sigma = \begin{bmatrix}
    \lambda_{\mathrm{lat}} & 0 \\ 0 & \lambda_{\mathrm{lon}} 
\end{bmatrix}
\]
While there are many methods for specifying nonstationary covariance functions \citep{nychka2002multiresolution,jun2008nonstationary,cressie2008fixed}, we follow the nonstationary specification provided by \cite{paciorek2006spatial},
\begin{equation}
\label{eq:nonstat_cov}
    C_n(\bm{s}_i, \bm{s}_j) = \phi |\Sigma_i|^{\frac{1}{4}}|\Sigma_j|^{\frac{1}{4}}\left|\frac{\Sigma_i + \Sigma_j}{2}\right|^{-\frac{1}{2}} \exp(-Q_{ij})
\end{equation}
where,
\[
Q_{ij} = (\bm{s}_i - \bm{s}_j)'\left(\frac{\Sigma_i + \Sigma_j}{2}\right)^{-1}(\bm{s}_i - \bm{s}_j)
\]
and,
\[
\Sigma_i = \begin{bmatrix}
    \lambda_{\mathrm{lat},i} & 0 \\ 0 & \lambda_{\mathrm{lon},i} 
\end{bmatrix}
\]
which serves as the nonstationary extension of the above stationary model using a squared exponential kernel.  We develop a validation methodology for estimating this nonstationary covariance function in Section \ref{sub:nonstation_val}.  As will be detailed in that section, our approach is to estimate the best \textit{local} stationary covariance at each held-out location in the validation set by maximizing the held-out likelihood for that location, and then rely on spatial smoothness assumptions to generalize to a nonstationary covariance model on the entire spatial domain.
\section{Estimation of spatially orthogonal factor loadings}
\label{sec:estimation}
% Estimation
% ---Explain probabilistic PCA enablement of a natural EM algorithm
% ---Show E-step
% ---Motivate that now U is orthogonal makes estimation subtle
% ---Assume spatial covariance is already specified; explain later
% ---Explain orthogonal procrustes problem has an analytic solution to tr(M'U) so can we use a minorization-maximization algorithm to find this solution
% ---Show M-step for U, explain M-steps for sigma and L are done using autograd.
Towards estimation, it is typically not possible to store $S$ nor $\Sigma$ in memory, and therefore it is not computationally feasible to perform an eigendecomposition of $S + \frac{1}{n}\Sigma$ as inferred in Theorem \ref{thm:MAP_U}. Instead, we use the latent variable construction from probabilistic PCA in addition to the prior distribution derived in Proposition \ref{prop:prior} and construct a minorization-maximization-within-EM (MM-EM) algorithm for estimating $U$.  Assuming the observations are independent conditioned on the latent state and the parameters, we have for a single observation, 
\[
p(\bm{y}_i|\bm{z}_i,U,L) = (2\pi\sigma^2)^{-\frac{m}{2}}\etr\left(-\frac{1}{2\sigma^2}(\bm{y}_i - UL\bm{z}_i)'(\bm{y}_i - UL\bm{z}_i)\right)
\]
The prior on $Z$ is an isotropic Gaussian; for each $i$th observation,
\[
p(\bm{z}_i) = (2\pi)^{-\frac{k}{2}}\etr\left(-\frac{1}{2}\bm{z}_i'\bm{z}_i\right)
\]
with the prior $p(U,L,\sigma^2|\Sigma)$ derived in the previous section. The $Q$ function is the posterior expectation with respect to the latent matrix $Z$ over the log posterior distribution. Let $\theta \coloneqq (U, L, \sigma^2)$,
\begin{align*}
    Q(\theta|\theta^{(t)}) &\overset{(c)}{=} \sum_{i = 1}^n -\frac{m}{2}\log(\sigma^2) - \frac{1}{2\sigma^2 }\left(\bm{y}_i'\bm{y}_i - \tr\left(2\bm{y}_i'UL\E[\bm{z}_i]\right) + \tr(\E[\bm{z}_i\bm{z}_i']L^2) \right)\\
    &\quad - \frac{1}{2}\tr(U \left((L^2 + \sigma^2 I_k)^{-1} - \sigma^{-2}I_k\right) U' \Sigma + \frac{1}{\sigma^2}\Sigma)\\
    &\quad + \sum_{j>i}^k \log\left((L_{jj}^2 + \sigma^2)^{-1} - (L_{ii}^2 + \sigma^2)^{-1}\right)
     + \sum_{j=1}^{k} (m-k)\log\left(((L_{jj}^2 + \sigma^2)^{-1} - \sigma^{-2}) \right)\\
     &\quad -\frac{1}{2}\left(\binom{m-k}{2}+3 \right)\log\left(\sigma^{2}\right)
     -\frac{3}{2} \sum_{j = 1}^k \log\left(L_{jj}^2 + \sigma^2\right)\\
     &\quad + \log\det(L)
\end{align*}
The posterior expectations are similar to the probabilistic PCA model,
\[
\E[\bm{z}_i] = (L^{(t)2} + \sigma^{2(t)} I_k)^{-1} L^{(t)} U^{(t)'}\bm{y}_i
\]
\[
\E[\bm{z}_i \bm{z}_i'] = \sigma^{2(t)} (L^{(t)2} + \sigma^{2(t)} I_k)^{-1} + \E[\bm{z}_i]\E[\bm{z}_i]'
\]
but easier to compute since the $U$ terms cancel and we only have to invert a diagonal matrix $(L^{(t)2} + \sigma^{2(t)} I_k)$. The superscript $(t)$ denotes the $t$th iteration of the EM algorithm.

\subsection{MM-algorithm for the M-step of spatially orthogonal loadings}
We use cyclic maximization steps for the three parameters $U,L,\sigma^2$ such that the maximization of each depends on the previous iteration of the other two.  For $L$ and $\sigma^2$ we rely on automatic gradient computation for the M-step given their dimensionality is small and the complexity is high for these terms. However, since $U$ is orthogonal, we derive a maximization solution directly.  Our observation is that the $Q$ function looks similar to an expression $\tr(M^T U)$, for some matrix $M$, which has a maximum for orthogonal $U$ known as the Procrustes solution \citep{schonemann1966generalized}.  When the objective function looks like $\tr(M^T U)$, if $M = B\Xi C'$ is the SVD of $M$, then the Procrustes solution is,
\[
BC' = \underset{U \in V_{k,m}}{\text{argmax}}\ \tr(M^T U)
\]
Taking an MM-EM approach, we utilize convexity properties to minorize the $Q$ function such that at each iteration of the minorization-maximization algorithm (which is employed within the EM iterations) we get an M-step of the form $\tr(M^T U)$ for some matrix $M$ that doesn't depend on $U$ \citep{hunter2004tutorial}. The M-step for $U$ requires maximizing,
\[
    Q(U|U^{(t)}) \overset{(c)}{=} \sum_{i = 1}^n \frac{1}{\sigma^2 }\tr\left(L\E[\bm{z}_i]\bm{y}_i'U\right) - \frac{1}{2}\tr( \left((L^2 + \sigma^2 I_k)^{-1} - \sigma^{-2}I_k\right) U' \Sigma U)
\]
where we have rearranged terms in the trace so that $U$ is on the right.  Before we can utilize Procrustes algorithm, we need to minorize the term from the prior to get rid of the quadratic dependence on $U$.
Start by noting that each diagonal element of $(L^2 + \sigma^2 I_k)^{-1} - \sigma^{-2}I_k$ is negative and define $\Lambda = -\left((L^2 + \sigma^2 I_k)^{-1} - \sigma^{-2}I_k\right)$, so the goal is minorization of $\tr( \Lambda U' \Sigma U)$. Now use the following relationship,
\begin{align*}
    \tr( \Lambda (U - U^{(s)})' \Sigma (U - U^{(s)})) &= \tr( \Lambda U' \Sigma U) - \tr( \Lambda U^{(s)'} \Sigma U) - \tr( \Lambda U' \Sigma U^{(s)}) + \tr( \Lambda U^{(s)'} \Sigma U^{(s)})
    % &= \tr( \Lambda U' \Sigma U) - \tr( \Lambda U^{(s)'} \Sigma U) - \tr( U^{(s)'} \Sigma U \Lambda) + \tr( \Lambda U^{(s)'} \Sigma U^{(s)})\\
    % &\overset{(c)}{=} \tr( \Lambda U' \Sigma U) - 2\tr( \Lambda U^{(s)'} \Sigma U)
\end{align*}
And rearranging so the term of interest is on the left,
\begin{align*}
\tr( \Lambda U' \Sigma U) &=  \tr( \Lambda (U - U^{(s)})' \Sigma (U - U^{(s)})) + 2\tr( \Lambda U^{(s)'} \Sigma U) - \tr( \Lambda U^{(s)'} \Sigma U^{(s)})\\
&\overset{(c)}{=}  \tr( \Lambda (U - U^{(s)})' \Sigma (U - U^{(s)})) + 2\tr( \Lambda U^{(s)'} \Sigma U)\\
&\overset{(c)}{\geq} 2\tr( \Lambda U^{(s)'} \Sigma U)
\end{align*}
Since $\tr( \Lambda (U - U^{(s)})' \Sigma (U - U^{(s)}))\geq 0$ and equal to $0$ when $U^{(s)}=U$, and $\tr( \Lambda U^{(s)'} \Sigma U^{(s)})$ does not depend on $U$, we have that $2\tr( \Lambda U^{(s)'} \Sigma U)$ is a valid minorization of $\tr( \Lambda U' \Sigma U)$. \cite{keys2016projection} develops a similar minorization for a different problem involving estimation of an orthogonal matrix. Plugging this minorization back into the Q function,
\begin{align*}
Q(U|U^{(t)},U^{(s)}) &\overset{(c)}{=} \sum_{i = 1}^n \frac{1}{\sigma^2 }\tr\left(L\E[\bm{z}_i]\bm{y}_i'U\right) - \frac{1}{2}\tr( \left((L^2 + \sigma^2 I_k)^{-1} - \sigma^{-2}I_k\right) U' \Sigma U)\\
&\overset{(c)}{\geq} \tr\left(\frac{1}{\sigma^2 } L\left(\sum_{i = 1}^n \E[\bm{z}_i]\bm{y}_i'\right)U\right) + \tr( \Lambda U^{(s)'} \Sigma U)\\
&\overset{(c)}{\geq} \tr\left(\left(\frac{1}{\sigma^2 } L\left(\sum_{i = 1}^n \E[\bm{z}_i]\bm{y}_i'\right) + \Lambda U^{(s)'} \Sigma\right) U \right)
\end{align*}
Where the $U^{(t)}$ parameter is implicitly conditioned on in the posterior expectations, and the $U^{(s)}$ is the previous iterate of $U$ within the minorization maximization algorithm.  Thus, the complete M-step with respect to $U$ is,
\begin{equation}
    \label{mstepU}
    U^{(s+1)} = B^{(s)}C^{(s)'}
\end{equation}
Where $B$ and $C$ are the orthogonal matrices from the SVD,
\[
\left(\frac{1}{\sigma^2 } L\left(\sum_{i = 1}^n \E[\bm{z}_i]\bm{y}_i'\right) + \Lambda U^{(s)'} \Sigma\right)' = B^{(s)}\Xi^{(s)} C^{(s)'}
\]
This completes our derivation of a minorization-maximization within EM algorithm for optimizing the $Q$ function with respect to the orthogonal loadings matrix $U$. However, this estimation is conditioned on $\Sigma$.  In the next section, we extend this MM-EM algorithm to include estimation of held-out spatial locations, which serves as the foundation of our validation approach to estimating the parameters within $\Sigma$.
\section{Methodology for spatial prior estimation}
\label{sec:validation}
% A validation approach for prior spatial covariance estimation
% ---Motivate that prior specification of Sigma is hard, affects the results greatly, and need a validation method for tuning its parameters
The parameterization of $\Sigma$ in the prior $p(U,L,\sigma^2 | \Sigma)$ can affect downstream inferences greatly. However, gradient methods for estimation of $\Sigma$ are not feasible since the normalizing constant is hard to compute and getting posterior samples of $U,L,\sigma^2$ is difficult in the settings we are considering. But due to its strong influence on the posterior, it behooves us to develop a computationally feasible methodology for estimating $\Sigma$.  We choose to use a validation framework for estimation.  

In this framework, spatial locations are selected at random and held-out from training.  From a modeling perspective, we still specify the loadings $U$ at all locations, but treat the held-out outcome variables as latent.  In this way, our MM-EM algorithm can be extended to handle those latent locations, yielding estimation of the loadings $U$ at all locations.  Predictions of the held-out outcome variables can then be assessed against the observed data and the validation loss can be used to choose the best parameterization of the spatial prior covariance $\Sigma$.

\subsection{MM-EM algorithm on held-out spatial locations}
% sub: Spatial block validation via EM
% ---Explain why MLE/MAP is not tractable
% ---Motivate validation as approximation of KL divergence between true distribution and model distribution
% ---Motivate spatial block validation
% ---Validation approach; can extend EM for missing locations while keeping dependence structure on latent U's
% ---Give some of the EM equations for this
% ---Motivate held-out MSE as loss function of interest, give heldout MSE Bayes estimator
Let $p$ be an indexed set of held-out spatial locations and let $\bm{y}_{i,p}$ be the corresponding vector of held-out (and consequently treated as latent) outcomes. Let $o$ be the indexed set of training locations and let $\bm{y}_{i,o}$ be the corresponding vector of training outcomes. We assume each outcome vector is ordered as $\bm{y}_i = (\bm{y}_{i,o}', \bm{y}_{i,p}')'$.

\subsubsection{Posterior distribution of latent states conditioned only on training data}
We wish to evaluate the posterior distribution $p(\bm{z}_i | \bm{y}_{o})$ which, as we will show, does not take the same form as $p(\bm{z}_i | \bm{y}_{i})$ when conditioned on all data, since $U_o$ is no longer orthogonal. Firstly, we have $p(\bm{z}_i | \bm{y}_{o}) = p(\bm{z}_i | \bm{y}_{i,o})$, since the $\bm{z}_i$ are independent across $i$, the $\bm{y}_i$ are conditionally independent across $i$ given the model, and the individual entries of $\bm{y}_o$ are conditionally independent across $i$ given the model. Implicitly conditioning on the parameters, 
\begin{align*}
p(\bm{z}_i | \bm{y}_{i,o}) &\propto \exp \left(-\frac{1}{2\sigma^2}(\bm{y}_{i,o} - U_o L \bm{z}_i)'(\bm{y}_{i,o} - U_o L \bm{z}_i)\right)\exp(-\frac{1}{2}\bm{z}_i'\bm{z}_i)\\
&\propto \exp \left(-\frac{1}{2\sigma^2}(-2\bm{z}_i' L U_o'\bm{y}_{i,o} + \bm{z}_i'LU_o'U_oL\bm{z}_i + \sigma^2 \bm{z}_i'\bm{z}_i)\right)\\
&\propto \exp \left(-\frac{1}{2\sigma^2}(-2\bm{z}_i' L U_o'\bm{y}_{i,o} + \bm{z}_i'(LU_o'U_oL + \sigma^2I_k)\bm{z}_i)\right)
\end{align*}
Thus, the posterior distribution,
\begin{equation}
\label{eq:cv_z_post}
\bm{z}_i | \bm{y}_{i,o} \sim \mathcal{N}(M^{-1}LU_o'\bm{y}_{i,o}, \sigma^2 M^{-1})    
\end{equation}
Where $M = (LU_o'U_oL + \sigma^2I_k)$.
\subsubsection{Expectation step in cross validation}
The posterior expectations of the log likelihood/prior(depending on which components are observed/missing, respectively) of observation $i$, amount to evaluating two different terms. With the expectation implicitly conditioned on $\bm{y}_{i,o}, U^{(t)},L^{(t)},\sigma^{2(t)}$, the two expectations of interest arise from the following terms within the $Q$ function,
\begin{align*}
    \E (\bm{y}_i - UL\bm{z}_i)'(\bm{y}_i - UL\bm{z}_i) &= \E (\bm{y}_i'\bm{y}_i - 2\bm{y}_i'UL\bm{z}_i + \bm{z}_i'L^2 \bm{z}_i)\\
    &= \E[\bm{y}_i'\bm{y}_i] - \E[2\bm{y}_i'UL\bm{z}_i] + \tr(\E[\bm{z}_i\bm{z}_i'] L^2)\\
    &= \bm{y}_{i,o}'\bm{y}_{i,o} + \E[\bm{y}_{i,p}'\bm{y}_{i,p}] - 2\bm{y}_{i,o}'U_{o}L\E[\bm{z}_i] - 2 \tr(\E[\bm{z}_i \bm{y}_{i,p}']U_p L) + \tr(\E[\bm{z}_i\bm{z}_i'] L^2)
\end{align*}
As such, the two expectations with respect to missing outcomes $\bm{y}_{i,p}$ are $\E[\bm{y}_{i,p}'\bm{y}_{i,p}|\bm{y}_{i,o}]$ and $\E[\bm{z}_i \bm{y}_{i,p}'|\bm{y}_{i,o}]$.

To evaluate $\E[\bm{y}_{i,p}'\bm{y}_{i,p}|\bm{y}_{i,o}] = \tr(\E[\bm{y}_{i,p}\bm{y}_{i,p}'|\bm{y}_{i,o}])$, we use the trace rotation and thus need to evaluate $\cov(\bm{y}_{i,p}|\bm{y}_{i,o})$ and $\E[\bm{y}_{i,p}|\bm{y}_{i,o}]$.
\begin{align*}
    \cov(\bm{y}_{i,p}|\bm{y}_{i,o}) &= \cov(U_p L \bm{z}_i + \sigma \bm{\epsilon}_{i,p}|\bm{y}_{i,o})\\
    &= U_pL\cov(\bm{z}_i|\bm{y}_{i,o})LU_p' + \sigma^2 I_p \\
    &= \sigma^2 U_pL(LU_o'U_oL + \sigma^2I_k)^{-1}LU_p' + \sigma^2 I_p 
\end{align*}
And,
\begin{align*}
    \E[\bm{y}_{i,p}|\bm{y}_{i,o}] &= \E[U_p L \bm{z}_i + \sigma^2 \bm{\epsilon}_{i,p}|\bm{y}_{i,o}]\\
    &= U_p L \E[\bm{z}_i|\bm{y}_{i,o}]\\
    &= U_p L M^{-1}LU_o'\bm{y}_{i,o}
\end{align*}
For the second expectation, note,
\begin{align*}
    \E_{\bm{z}_i,\bm{y}_{i,p}|\bm{y}_{i,o}}[\bm{z}_i \bm{y}_{i,p}'] &= \E_{\bm{z}_i|\bm{y}_{i,o}}[\bm{z}_i \E_{\bm{y}_{i,p}|\bm{z}_i,\bm{y}_{i,o}}[\bm{y}_{i,p}'|\bm{z}_i]]\\
    &= \E_{\bm{z}_i|\bm{y}_{i,o}}[\bm{z}_i \E_{\bm{y}_{i,p}|\bm{z}_i}[\bm{y}_{i,p}'|\bm{z}_i]]
\end{align*}
The second equality holds by the conditional independence assumption in the model, $y_{i,j} \perp y_{i,j^*} | U,L,\sigma^2,\bm{z}_i$ for $j\ne j^*$. The inner expectation is the assumed mean function of the likelihood, $\E_{\bm{y}_{i,p}|\bm{z}_i}[\bm{y}_{i,p}'|\bm{z}_i] = (U_p^{(t)} L^{(t)} \bm{z}_i)'$. Plugging this back into the expectation of interest,
\begin{equation}
\label{eq:cv_2ndEstep}
\E_{\bm{z}_i,\bm{y}_{i,p}|\bm{y}_{i,o}}[\bm{z}_i \bm{y}_{i,p}'] = \E_{\bm{z}_i|\bm{y}_{i,o}}[\bm{z}_i \bm{z}_i'] L^{(t)} U^{(t)'}_p
\end{equation}
Plugging these back into the $Q$ function, we have,
\begin{align*}
    Q(\theta|\theta^{(t)}) &\overset{(c)}{=} \sum_{i = 1}^n -\frac{m}{2}\log(\sigma^2) - \frac{1}{2\sigma^2 }\bigg(\bm{y}_{i,o}'\bm{y}_{i,o} + \E[\bm{y}_{i,p}'\bm{y}_{i,p}] - 2\bm{y}_{i,o}'U_{o}L\E[\bm{z}_i] - 2 \tr(\E[\bm{z}_i \bm{y}_{i,p}']U_p L)\\ 
    &\quad + \tr(\E[\bm{z}_i\bm{z}_i']L^2)\bigg) +\log \pi(U,L,\sigma^2|\Sigma)\\
    &\overset{(c)}{=} \sum_{i = 1}^n -\frac{m}{2}\log(\sigma^2) - \frac{1}{2\sigma^2 }\bigg(\bm{y}_{i,o}'\bm{y}_{i,o} + \tr\left(\sigma^2 U_pL(LU_o'U_oL + \sigma^2I_k)^{-1}LU_p' + \sigma^2 I_p\right)\\ 
    &\quad + \tr(U_p L M^{-1}LU_o'\bm{y}_{i,o} \bm{y}_{i,o}'U_o L M^{-1} L U_p')\\ 
    &\quad - 2\bm{y}_{i,o}'U_{o}L\E[\bm{z}_i] - 2 \tr(\E[\bm{z}_i \bm{z}_i'] L^{(t)} U^{(t)'}_p U_p L) + \tr(\E[\bm{z}_i\bm{z}_i']L^2)\bigg)\\
    &\quad +\log \pi(U,L,\sigma^2|\Sigma)
\end{align*}

\subsubsection{Maximization step for U in cross validation}
In the case when there are held-out data, the M-step with respect to $U$ is changed by these new expectations in the $Q$ function,
\begin{align*}
&\frac{1}{\sigma^2 }\bigg(\tr(L\sum_{i=1}^n\left(\E[\bm{z}_i]\bm{y}_{i,o}'\right)U_{o}) + \tr(L\sum_{i=1}^n\left(\E[\bm{z}_i \bm{y}_{i,p}']\right)U_p)\bigg)\\ 
&=\\ 
&\frac{1}{\sigma^2 }\bigg(\tr(L\begin{bmatrix}
    \sum_{i=1}^n\left(\E[\bm{z}_i]\bm{y}_{i,o}'\right) & \sum_{i=1}^n\left(\E[\bm{z}_i \bm{y}_{i,p}']\right)
\end{bmatrix}U) \bigg)\\
&=\\
&\frac{1}{\sigma^2 }\bigg(\tr(L\begin{bmatrix}
    \sum_{i=1}^n\left(\E[\bm{z}_i]\bm{y}_{i,o}'\right) & \sum_{i=1}^n\left(\E[\bm{z}_i \bm{z}_i']\right)L^{(t)} U^{(t)'}_p
\end{bmatrix}U) \bigg)
\end{align*}
Where $U = \begin{bmatrix} U'_o & U'_p \end{bmatrix}'$. Putting this together, we have the updated M-step for $U$ is,
\begin{equation*}
    U^{(s+1)} = B^{(s)}C^{(s)'}
\end{equation*}
Where $B$ and $C$ are the orthogonal matrices from the SVD,
\[
\left(\frac{1}{\sigma^2 } L\begin{bmatrix}
    \sum_{i=1}^n\left(\E[\bm{z}_i]\bm{y}_{i,o}'\right) & \sum_{i=1}^n \E[\bm{z}_i \bm{z}_i']L^{(t)} U^{(t)'}_p
\end{bmatrix} + \Lambda U^{(s)'} \Sigma\right)' = B^{(s)}\Xi^{(s)} C^{(s)'}
\]
It is worth taking a moment to discuss the difference between the $t$th iteration of the EM algorithm and the $s$th iteration of the minorization-maximization algorithm within the EM algorithm.  During the MM-EM algorithm, an E-step is performed conditioned on the value $U^{(t)}$. Then at the M-step, we use a minorization of the $Q$ function at the $t$th iteration.  As such, since the term $U^{(t)}$ within the M-step equation is from the E-step, it does not update during the minorization-maximization steps. The only $U^{(s)}$ term from the minorization of the $Q$ function occurs within the quadratic term in the prior. Only this term from the prior is updated during the minorization-maximization step nested within each EM step.

\subsubsection{Prediction of held-out data}
\label{subsub:heldout_likelihood}
We use the log-likelihood on held-out data as our validation metric of interest. The predictive distribution of the held-out outcomes $\bm{y}_{i,p}$ conditioned on the observed training outcomes $\bm{y}_{i,o}$ is obtained by marginalizing over the latent states $\bm{z}_i$.

Since we cannot marginalize over the full posterior of $U,L,\sigma^2$, we approximate the held-out likelihood by conditioning on the local MAP estimates $(U^\dagger,L^\dagger,\sigma^{2\dagger})$ obtained for each spatial prior $\Sigma$ during the validation process. Let $\hat{\bm{y}}_{i,p} = \E[\bm{y}_{i,p}| U^\dagger,L^\dagger,\sigma^{2\dagger}, \bm{y}_{i,o}]$ be the predictive mean and let $V_p = \cov(\bm{y}_{i,p}| U^\dagger,L^\dagger,\sigma^{2\dagger}, \bm{y}_{i,o})$ be the predictive covariance matrix.  The loss function with respect to the parameters of $\Sigma$ is,
\[
\mathcal{L}(\Sigma) = \sum_{i = 1}^n \left( -\frac{1}{2} \log |V_{p,\Sigma}| - \frac{1}{2} (\bm{y}_{i,p} - \hat{\bm{y}}_{i,p})' V_{p,\Sigma}^{-1} (\bm{y}_{i,p} - \hat{\bm{y}}_{i,p}) \right)
\]
Where $\Sigma$ depends on a sill parameter $\phi$ and length scale parameters $(\bm{\lambda}_{\mathrm{lat}},\bm{\lambda}_{\mathrm{lon}})$ as specified in Section \ref{sub:spatial_cov}.
% \subsubsection{Prediction of held-out data}
% We use mean squared error (MSE) on held-out data as our loss function of interest.  The Bayes estimator for this loss function is the posterior expectation of the latent outcomes conditioned on observed outcomes $\E[\bm{y}_p|\bm{y}_o]$. Of course, we cannot marginalize over the full posterior of $U,L,\sigma^2$ to evaluate this expectation, and we approximate it by conditioning on the local MAP estimates $\E[\bm{y}_p| U^\dagger,L^\dagger,\sigma^{2\dagger}, \bm{y}_o]$. The combination of length scale and variance parameters of $\Sigma$ that yields the lowest MSE is our estimator of $\Sigma$,
% \[
% \mathcal{L}(\Sigma) = \frac{1}{m_p n} \sum_{i = 1}^n ||\bm{y}_{i,p} - \E[\bm{y}_{i,p}| U_{\Sigma}^\dagger,L_{\Sigma}^\dagger,\sigma_{\Sigma}^{2\dagger}, \bm{y}_{i,o}]||_2^2
% \]
% Where $m_p$ is the number of locations in the held-out set $p$. 

\subsection{Validation approach for nonstationary covariance estimation}
\label{sub:nonstation_val}
Now that an MM-EM algorithm has been developed for fitting the model when a set of locations has been held-out, we develop an algorithm using validation for estimating the prior nonstationary spatial covariance matrix. Our objective is to estimate the sill parameter $\phi$ and the length scales $\{\lambda_{\mathrm{lat},i},\lambda_{\mathrm{lon},i}\}_{i = 1}^m$ for each of the $m$ spatial locations. To that end, we randomly choose a held-out set of spatial locations $\mathcal{D}_{\mathrm{hold}}$ from the spatial domain $\mathcal{D}$.

To estimate the sill $\phi$, we start with an initial stationary length scale $(\lambda_{\mathrm{lat,init}},\lambda_{\mathrm{lon,init}})$ used to construct a stationary squared exponential correlation matrix $C_{s,\lambda_{\mathrm{lat,init}},\lambda_{\mathrm{lon,init}}}$ for the spatial domain $\mathcal{D}$. Then a set of sill values $\phi \in \bm{p}$ are each used to fit an SOFM, and the MAP parameters are then used to compute the held-out likelihood from Section \ref{subsub:heldout_likelihood}.  The $\phi$ with the largest corresponding held-out likelihood is our estimator.

After the sill is estimated, we need to estimate $(\lambda_{\mathrm{lat},s},\lambda_{\mathrm{lon},s})$ for each spatial location $s \in \mathcal{D}$.  We do this in two stages. In the first stage, we create a stationary covariance matrix for each pair of length scales in a specified set $(\lambda_{\mathrm{lat}},\lambda_{\mathrm{lon}})\in (\bm{\ell}\times \bm{\ell})$. The SOFM model is fit for each pair of length scales, and the held-out likelihood is computed and stored for each held-out spatial location $s\in\mathcal{D}_{\mathrm{hold}}$. The $(\lambda_{\mathrm{lat},s},\lambda_{\mathrm{lon},s})$ that maximize the held-out likelihood for each location $s\in\mathcal{D}_{\mathrm{hold}}$ are selected. Then, to generalize our estimates to the entire spatial domain $\mathcal{D}$, we assume the length-scales follow a smooth spatial surface and fit that surface in a second stage of estimation.

\subsubsection{P-splines for smoothing local length scales}
We use the two-dimensional P-spline model, which we briefly cover here and refer readers to \cite{eilers2003multivariate} for more information. The log of the length scales is modeled to ensure positivity of our estimates. Since the two vectors of log length scale estimates are evaluated at the same locations $\mathcal{D}_{\mathrm{hold}}$, the following B-spline matrices can be used to fit both of them.  Let $B_y \in \R^{h \times r_y}$ be a B-spline basis matrix for the y-axis evaluated at $r_y$ knots and $B_x \in \R^{h \times r_x}$ be a B-spline basis matrix for the x-axis evaluated at $r_x$ knots.  The two dimensional B-spline basis matrix $B$ combines $B_y$ and $B_x$ as follows,
\[
B \coloneqq (B_x \otimes 1_{r_y}') \odot (1_{r_x}' \otimes B_{y})
\]
The spline regression is then regularized using second-order difference matrices $D_x$ and $D_y$. These difference matrices are constructed such that the spline parameters for neighboring knots are penalized with respect to their difference.  Please see \cite{eilers2003multivariate} for details. In the two-dimensional case, the difference matrices are extended to two-dimensions by using $P_x \coloneqq (D_x'D_x \otimes I_{r_y})$ and $P_y \coloneqq (I_{r_x} \otimes D_y'D_y)$. The following loss functions are used to fit the P-splines,
\[
\mathcal{L}(\bm{\alpha}_\mathrm{lat}; \hat{\bm{\lambda}}_\mathrm{lat},\gamma_x, \gamma_y) = ||\log(\hat{\bm{\lambda}}_{\mathrm{lat}}) - B \bm{\alpha}_\mathrm{lat}||_2^2 + \gamma_x \bm{\alpha}_\mathrm{lat}'P_x\bm{\alpha}_\mathrm{lat} + \gamma_y \bm{\alpha}_\mathrm{lat}'P_y\bm{\alpha}_\mathrm{lat}
\]
with a similar loss for the longitudinal length scales,
% \[
% \mathcal{L}(\bm{\alpha}_\mathrm{lat}; \hat{\bm{\lambda}}_\mathrm{lat},\gamma_x, \gamma_y) = ||\log(\hat{\bm{\lambda}}_{\mathrm{lat}}) - B \bm{\alpha}_\mathrm{lat}||_2^2 + \gamma_x ||P_x\bm{\alpha}_\mathrm{lat}||_2^2 + \gamma_y ||P_y\bm{\alpha}_\mathrm{lat}||_2^2
% \]
which has the following minimizer,
\[
\hat{\bm{\alpha}}_\mathrm{lat} = (B'B + \gamma_x P_x + \gamma_y P_y)^{-1}B'\log(\hat{\bm{\lambda}}_\mathrm{lat})
\]
% \[
% \hat{\bm{\alpha}}_\mathrm{lat} = (B'B + \gamma_xP_x'P_x + \gamma_yP_y'P_y)^{-1}B'\log(\hat{\bm{\lambda}}_\mathrm{lat})
% \]
% and
% \[
% \hat{\bm{\alpha}}_\mathrm{lon} = (B'B + \gamma_xP_x'P_x + \gamma_yP_y'P_y)^{-1}B'\log(\hat{\bm{\lambda}}_\mathrm{lon})
% \]
In this work, we set the number of knots to $r_x = r_y = 10$, and use generalized cross-validation to estimate the $\gamma$-regularization parameters \citep{craven1978smoothing,eilers1996flexible}, which are estimated separately for the latitudinal and longitudinal models.

Once these $\alpha$ parameters are fit, we get predictions for the latitudinal and longitudinal length scales for every location by reevaluating the B-spline basis functions $\Tilde{B}$ at all locations in $\mathcal{D}$, for the same knots that were used during training and then plugging $\Tilde{\lambda} = \exp(\Tilde{B}\hat{\bm{\alpha}})$. The details are summarized in Algorithm \ref{alg:nonstat_val}.

\begin{algorithm}
\caption{Nonstationary spatial model fitting}
\label{alg:nonstat_val}
\begin{algorithmic}
\State \textbf{Input:} Spatial domain $\mathcal{D}$, discretized length scale domain $\bm{\ell} \coloneqq \{\ell_i\}_{i}$ and discretized sill domain $\bm{p} \coloneqq \{p_i\}_{i}$.
\State \textbf{Step 1: Selection of hold-out set from spatial domain} 
\State Randomly select set of held-out spatial locations $\mathcal{D}_{\mathrm{hold}}$ from spatial domain $\mathcal{D}$. 
\State Define $\mathcal{D}_{\mathrm{keep}} = \mathcal{D}\cap \mathcal{D}_{\mathrm{hold}}^C$
\State \textbf{Step 2: Sill estimation}
\State Conditioned on an initial $(\lambda_{\mathrm{lat},\mathrm{init}}, \lambda_{\mathrm{lon},\mathrm{init}})$, create stationary correlation matrix $C_{s,\mathrm{init}}$ using stationary kernel $C_s$ from Section \ref{sub:spatial_cov}.
\For{each sill $\phi \in \bm{p}$}
    \State Run $\mathrm{SOFM}(\mathcal{D}_{\mathrm{keep}},\mathcal{D}_{\mathrm{hold}},\Sigma_{s,\phi})$ conditioned on $\Sigma_{s,\phi} = \phi \cdot C_{s,\mathrm{init}}$
    \State Compute and store held-out likelihood for $\mathcal{D}_{\mathrm{hold}}$
\EndFor
\State Choose $\hat{\phi} \in \bm{p}$ that maximizes held-out likelihood over $\mathcal{D}_{\mathrm{hold}}$
\State \textbf{Step 3: Local covariance estimation}
\For{each pair $(\lambda_{\mathrm{lat}},\lambda_{\mathrm{lon}})\in (\bm{\ell}\times \bm{\ell})$}
\State Create stationary covariance matrix $\Sigma_{s,\lambda_{\mathrm{lat}},\lambda_{\mathrm{lon}}} = \hat{\phi} \cdot C_{s,\lambda_{\mathrm{lat}},\lambda_{\mathrm{lon}}}$
\State Run $\mathrm{SOFM}(\mathcal{D}_{\mathrm{keep}},\mathcal{D}_{\mathrm{hold}},\Sigma_{s,\lambda_{\mathrm{lat}},\lambda_{\mathrm{lon}}})$
\State Compute and store held-out likelihood for each held-out location $d \in \mathcal{D}_{hold}$
\EndFor
\For{each held-out location $d \in \mathcal{D}_{hold}$}
    \State Choose pair $(\hat{\lambda}_{\mathrm{lat},d},\hat{\lambda}_{\mathrm{lon},d})\in (\bm{\ell}\times \bm{\ell})$ that maximizes held-out likelihood for $d$th location.
\EndFor
\State \textbf{Step 4: Two-dimensional spline smoothing for generalization}
\State Fit a two-dimensional P-spline to the log of $\hat{\bm{\lambda}}_{\mathrm{lat}}$.
\State Fit a two-dimensional P-spline to the log of $\hat{\bm{\lambda}}_{\mathrm{lon}}$.
\State \textbf{Step 5: Prediction on $\mathcal{D}$}
\State Use the two fitted P-splines to predict length scale parameters $\Tilde{\bm{\lambda}}_{\mathrm{lat}}$, $\Tilde{\bm{\lambda}}_{\mathrm{lon}}$ for all locations in $\mathcal{D}$.
\State \textbf{Output:} The full nonstationary covariance function evaluated at all locations in $\mathcal{D}$.
\end{algorithmic}
\end{algorithm}

% \subsection{Validation approach for stationary covariance estimation}
% The validation approach for stationary prior covariance estimation is similar to the approach for nonstationary prior covariance estimation, except that only a single set of length scales is estimated for the entire domain.  As such Algorithm \ref{alg:nonstat_val} is modified slightly to remove local selection of the best parameters followed by smoothing.  The details are available in Algorithm \ref{alg:stat_val}.
% \begin{algorithm}
% \caption{Stationary spatial model fitting}
% \label{alg:stat_val}
% \begin{algorithmic}
% \State \textbf{Input:} Spatial domain $\mathcal{D}$, discretized length scale domain $\bm{\ell} \coloneqq \{\ell_i\}_{i}$.
% \State \textbf{Step 1: Selection of hold-out set from spatial domain} 
% \State Choose hold-out set $\mathcal{B}_{\mathrm{hold}}$ from spatial domain $\mathcal{D}$. 
% \State \textbf{Step 2: Selection of best length scales}
% \For{each pair of length scales in $(\bm{\ell}\times \bm{\ell})$}
%     \State Evaluate the held-out MSE over all blocks $b\in \mathcal{B}_{\mathrm{hold}}$
% \EndFor
% \State Select $\hat{\lambda}_{\mathrm{lat}}$ and $\hat{\lambda}_{\mathrm{lon}}$ that minimize held-out MSE
% \State \textbf{Output:} The stationary covariance function.
% \end{algorithmic}
% \end{algorithm}

\section{Simulation study}
\label{sec:synth}
The simulation study is aimed at demonstrating the empirical performance of the nonstationary model and methodology from Algorithm \ref{alg:nonstat_val}. The goal is to study estimation of $U,L,\sigma^2$. We run our study on three different data generating spatial covariance models: a stationary model, a nonstationary model, where the length scales evolve smoothly over the spatial domain, and a shape model, where there is a circle in the center of the spatial domain with long length scales and short length scales outside of the circle.
%the study setup
We study two spatial domains of $400$ and $2500$ spatial locations, along with sample sizes $(200, 500, 1000, 5000, 10000)$. There are two $\sigma^2 \in \{5,10\}$. In each simulation, the true $U$ matrix is chosen to be the first $k$ eigenvectors of the underlying spatial covariance, the true eigenvalue parameters $L$ are chosen to maximize the prior conditioned on $U$, and then the synthetic data are simulated according to the likelihood model. We take $k=3$ in this study. Finally, there are twenty replicate simulations. For model setting, we consider maximum spatial lags of $2$ and $20$, initial length scales $\lambda_i \in \{1,3,5\}$ and sill values $\phi \in \{1,5,10,50,100,500,1000,5000,10000,50000\}$.

\begin{figure}
    \centering
    % \begin{subfigure}{\linewidth}
        \begin{subfigure}{0.74\linewidth}
            \includegraphics[width=\linewidth]{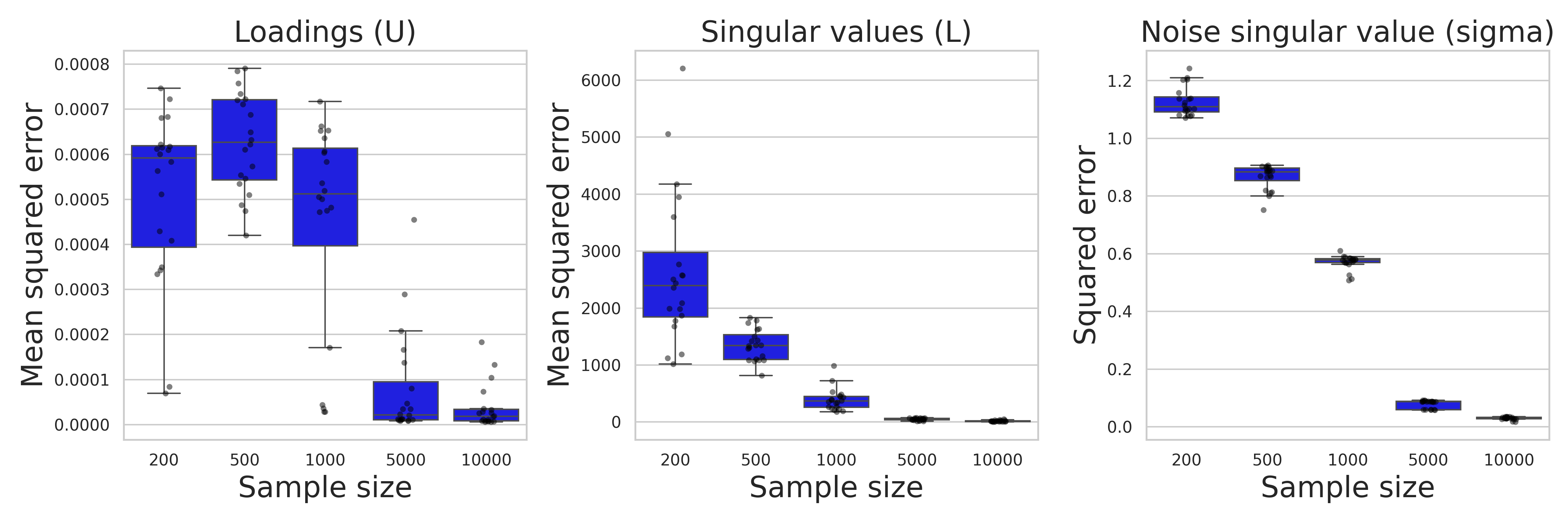} 
        \end{subfigure}
        \begin{subfigure}{0.25\linewidth}
            \includegraphics[width=\linewidth]{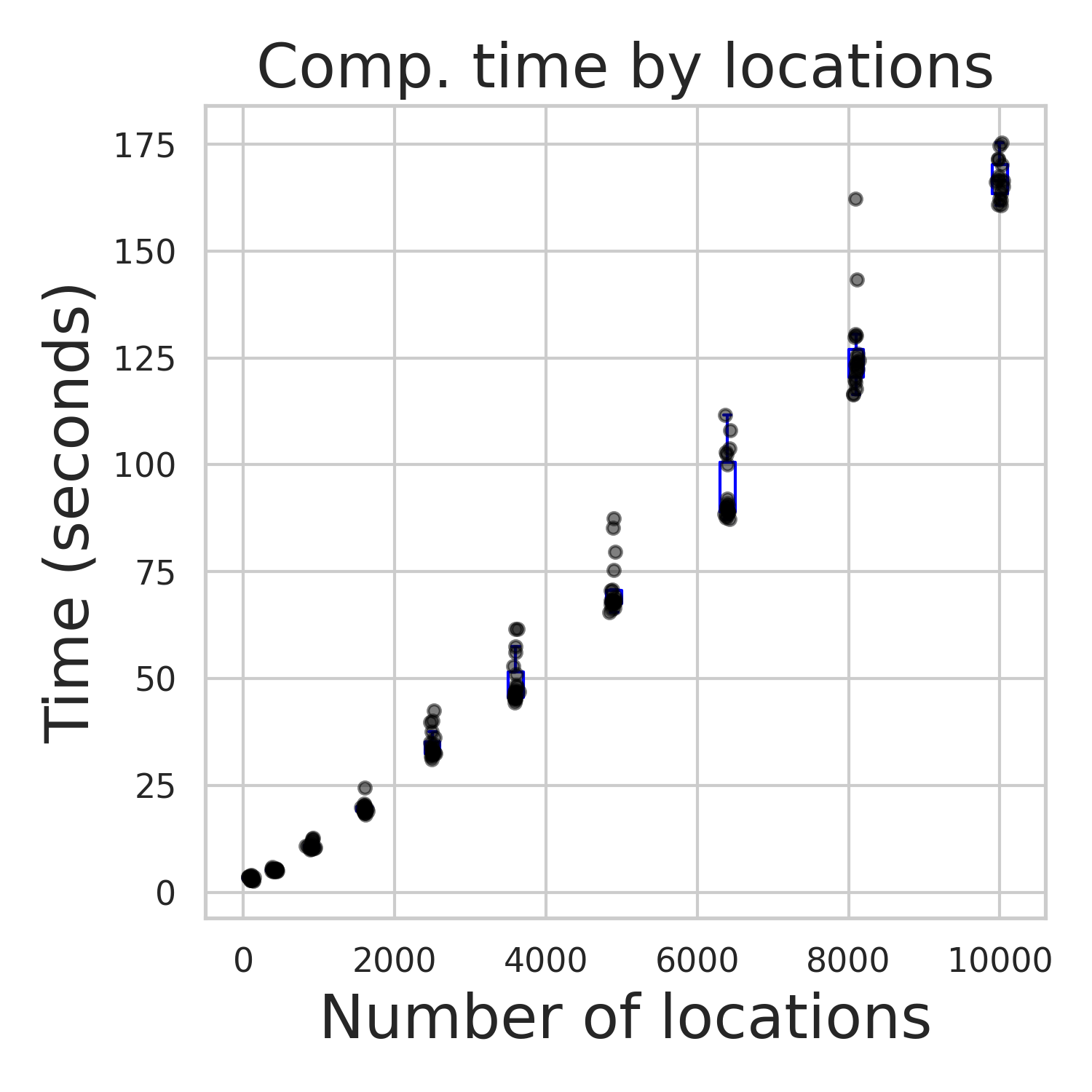}
        \end{subfigure}
    \caption{(\textit{Left}) SOFM run on synthetic data generated as described in Section \ref{sec:synth}. For each dataset, mean squared error is assessed for the $U,L,\sigma^2$ parameters. (\textit{Right}) A synthetic study was run on a single core to assess computational time against number of spatial locations, demonstrating linear computational complexity. Study was run with 3000 samples/features per dataset and $k=3$ factors.}
    \label{fig:main_synth}
\end{figure}

%the results of the parameter estimates
Please see Figure \ref{fig:main_synth} for results of the studies. There are two main approximations made in the Algorithm \ref{alg:nonstat_val} that may affect the study results. The first is that the prior variance and prior length scales are optimized in two separate conjugate steps and only once in total. The second is that the maximum spatial lag considered in the prior covariance is truncated for computation purposes. Results from Figure \ref{fig:main_synth} demonstrate that estimation of $U,L,\sigma^2$ is not sensitive to these approximations.
Appendix \ref{sec:app_synth} contains additional studies for varying values of $\sigma$, spatial domain size, as well as different maximum spatial lags of the prior covariance function.

\section{Case Study: Spatial transcriptomics from the DLPFC brain region}
\label{sec:DLPFC}
The barcoded-array Spatial Transcriptomics technology was introduced by \cite{staahl2016visualization}, and was subsequently extended and commercialized through technological platforms including 10x Genomics Visium. Spatial transcriptomics measures gene expression, i.e.\ the relative abundance of RNA transcripts associated with genes, across up to tens of thousands of genes while retaining the spatial coordinates associated with each measurement location within a tissue section. These technologies complement earlier advances in single-cell RNA sequencing (scRNA-seq), which characterize gene expression in individual cells but generally do not retain information about the cells' spatial locations within the tissue \citep{amezquita2020orchestrating}. By preserving this spatial context, spatial transcriptomics can be used to study tissue architecture and the spatial organization of cell types, cell states, and gene expression programs \citep{rao2021exploring}.

The dorsolateral prefrontal cortex (DLPFC) dataset used here was introduced by \cite{maynard2021transcriptome} and consists of spatially resolved gene expression measurements from postmortem human brain tissue sections from the DLPFC region. The full study includes 12 tissue sections from 3 donors, with manual annotations of the cortical layers and white matter providing an anatomical reference for evaluating spatial gene expression patterns. The data were generated using the 10x Genomics Visium platform, which measures gene expression at spatially indexed capture locations (referred to as spots) approximately 55 $\mu$m in diameter and spaced approximately 100 $\mu$m center-to-center, with up to approximately 5,000 spots per tissue section.

We analyzed section 151675 from this dataset, which contains expression measurements for 3,592 in-tissue Visium spots and 33,538 gene features. The input expression matrix consists of expression counts per gene and spot, which were normalized and log-transformed following workflows described by \citet{crowell2025orchestrating,amezquita2020orchestrating}, and subsequently mean-centered per gene across spots. We performed three sets of analyses: a filtered analysis with low-expression filtering applied to retain genes with positive expression in at least 20 spots (consistent with the filtering threshold used by \citet{shang2022spatially}), a full-gene analysis retaining all 33,538 genes, and a `highly variable gene' (HVG) analysis selecting the top 3,000 HVGs following \citet{crowell2025orchestrating,amezquita2020orchestrating}. The low-expression filtering retains 14,638 genes. Most spatial transcriptomics analysis workflows include low-expression filtering due to high sparsity and large numbers of genes with very low expression in the raw measurements; these genes provide little biological information and modeling them is computationally challenging \citep{crowell2025orchestrating,amezquita2020orchestrating}. We treat the filtered analysis as our main analysis in the results below, and include the full-gene analysis in the appendix to show the advantage of our spatial model both computationally as well as its ability to regularize in high dimensions. HVGs are genes with the highest expression variability across cells or spots, and are typically selected as a preprocessing step during workflows to focus on biologically relevant genes that contribute strongly to expression heterogeneity, and to reduce the number of input genes for computational purposes \citep{amezquita2020orchestrating}. However, conventional HVG selection does not incorporate spatial information and may remove some genes with spatial expression patterns \citep{weber2023nnsvg}; hence we performed all three analyses for demonstration purposes.

We note that in many common workflows for single-cell and spatial transcriptomics and existing factor-analysis methods, expression counts are normalized and subsequently transformed via $f(x_{ij}) = \log(x_{ij}+1)$, and the resulting transformed values are modeled as continuous variables \citep{shang2022spatially,stuart2019comprehensive,wolf2018scanpy,palla2022squidpy,amezquita2020orchestrating,crowell2025orchestrating}. Although this Gaussian modeling assumption is an approximation \citep{townes2019feature}, we adopt it here to enable direct comparison with PCA-based methods and to facilitate computationally efficient estimation, allowing the methodological development to focus on extending spatial covariance modeling from the stationary to the nonstationary setting.

\subsection{How does SOFM fit into spatial transcriptomics analysis workflows?}

Spatial transcriptomics analysis workflows typically begin with a gene expression count matrix measured at a set of spatial locations, together with spatial coordinates and any associated tissue images. Initial analysis steps include quality control, filtering, normalization, log-transformation, and feature selection. This is followed by construction of a low-dimensional representation of the gene expression matrix. In many standard workflows, this dimensionality reduction step is performed using PCA \citep{crowell2025orchestrating,palla2022squidpy}, consistent with established workflows for scRNA-seq data \citep{stuart2019comprehensive,wolf2018scanpy,amezquita2020orchestrating}; the PCA calculation itself does not use spatial coordinates. Alternatively, for spatial transcriptomics data, the dimensionality reduction step may be adapted or replaced using methods that take spatial coordinates into account, so that spatial information is incorporated into the resulting low-dimensional representation \citep{crowell2025orchestrating}.

The resulting top principal components (PCs) or other low-dimensional representations may then be used as the input for further downstream analyses, such as nearest-neighbor graph construction, clustering and spatial domain detection, interpretation of genes or gene sets, visualization, and other analyses. The corresponding gene coefficients also provide a gene-level view, allowing each spatial factor to be interpreted through genes with large positive or negative coefficients along that factor. The key distinguishing feature of SOFM is that it incorporates spatial information while preserving orthogonality of the inferred spatial patterns, thereby reducing redundancy among the spatial components.

\subsubsection{Comparison to other dimensionality reduction models}
Several existing methods have proposed modified dimensionality reduction models to incorporate spatial information. Examples include SpatialPCA \citep{shang2022spatially}, MEFISTO \cite{velten2022mefisto}, nonnegative spatial factorization \citep{townes2023nonnegative}, and GraphPCA \citep{yang2024graphpca}. SpatialPCA is a particularly relevant comparison, which builds on probabilistic PCA by incorporating a prior spatial covariance on the latent factors. SOFM fits into the same broad workflow step, as a spatially aware dimensionality reduction method after preprocessing and before downstream analyses. 

There are three main differences between SOFM and SpatialPCA. First, SpatialPCA uses a stationary spatial prior covariance, whereas SOFM estimates a location-varying nonstationary spatial prior covariance. Second, SpatialPCA uses a sample-size-dependent plug-in bandwidth selector, i.e.\ the Sheather-Jones method \citep{sheather1991sheatherjones} for datasets with at most 5,000 spatial locations and Silverman's rule-of-thumb \citep{silverman2018density} for larger datasets, whereas SOFM selects covariance parameters by maximizing held-out predictive likelihood. Third, SOFM achieves linear computational complexity in the number of spatial locations, holding other model dimensions fixed. The dense SpatialPCA implementation includes quadratic operations in the number of spatial locations, although low-rank approximations and a sparse-kernel option are also available. Computational complexity with respect to the number of spatial locations is important in this research regime, and methods have been proposed with the goal of linear computational complexity \citep{weber2023nnsvg}. 

\subsection{Spatially orthogonal factors in the DLPFC}
One defining feature of the DLPFC dataset provided by \cite{maynard2021transcriptome} is the joint availability of manual annotations of the cortical layers and white matter, together with the spatial transcriptomics measurements. This additional information about the structure of the tissue has enabled researchers to analyze gene expression patterns within the laminar structure as well as to identify non-laminar spatial patterns. We build on the analyses of \cite{maynard2021transcriptome} by developing a supervised laminae-specific PCA-like framework that conditions on the laminar structure, and use SOFM for studying general (i.e.\ potentially non-laminar) spatial modes of gene expression covariation.

\subsubsection{Modes of gene expression covariation within the laminar structure}
We extend \cite{maynard2021transcriptome}'s analysis to characterize gene expression coefficients associated with the cortical layers by developing a model that conditions on the layers as the spatial loadings. Specifically, we develop an extension of the probabilistic PCA method where the spatial loadings $U$ are fixed to an orthogonalization of the laminar structure, obtained by normalizing the columns of the layer-indicator matrix, and the eigenvalue parameters $L$ and $\sigma^2$ are estimated using EM. We call this model Laminae-PCA, and use it as a supervised anatomical reference for comparison with SOFM. Please see Appendix Section \ref{sec:laminae_pca} for more details on the model and methodology.

\subsubsection{SOFM: General (potentially non-laminar) modes of gene expression covariation}
In addition to laminar results, \cite{maynard2021transcriptome} identified non-laminar spatially variable genes, including \textit{HBB}, \textit{IGKC}, and \textit{NPY}, with expression patterns associated with blood, immune, and neuronal cell populations, respectively, using methods to identify spatially variable genes \citep{svensson2018spatialde}. We continue this line of scientific reasoning using SOFM and examine whether genes highlighted in the original study receive large-magnitude coefficients in the SOFM decomposition. In particular, we find that \textit{HBB} and \textit{IGKC}, which are associated with blood and immune cell populations, respectively, occur among the genes with the most extreme coefficients in one or more SOFM factors.

\subsubsection{Results}
\begin{figure}
    \centering
    \begin{subfigure}{0.475\linewidth}
        \centering
        \includegraphics[width=\linewidth]{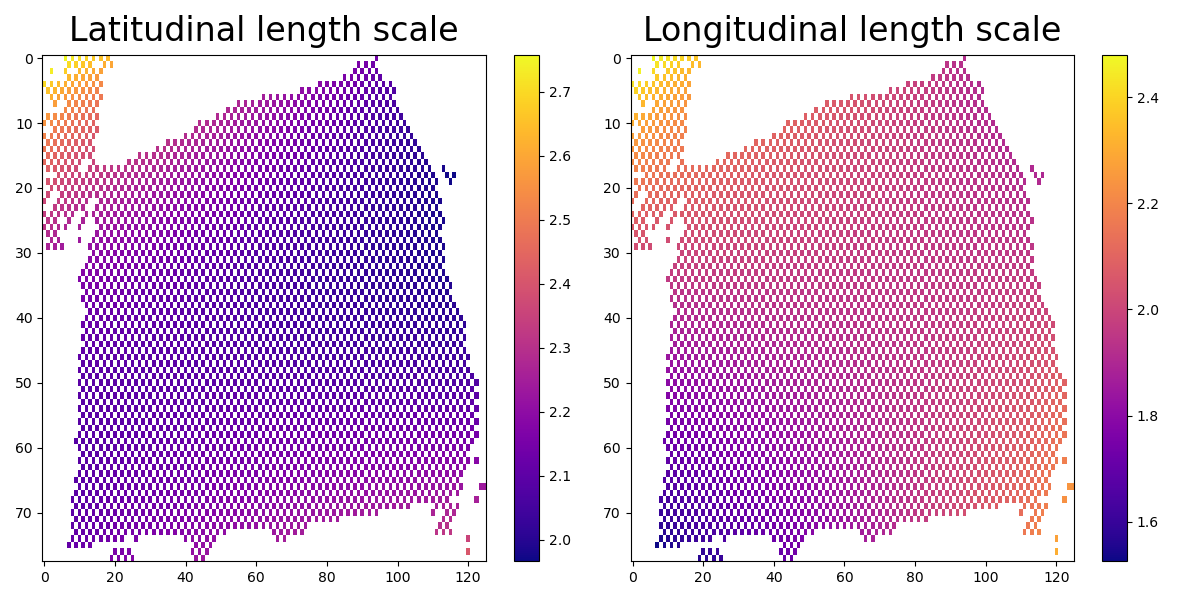}    
    \end{subfigure}
    \begin{subfigure}{0.25\linewidth}
        \centering
        \includegraphics[width=\linewidth]{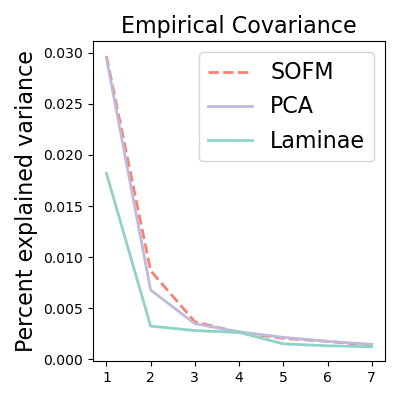}
    \end{subfigure}
    \begin{subfigure}{0.25\linewidth}
        \centering
        \includegraphics[width=\linewidth]{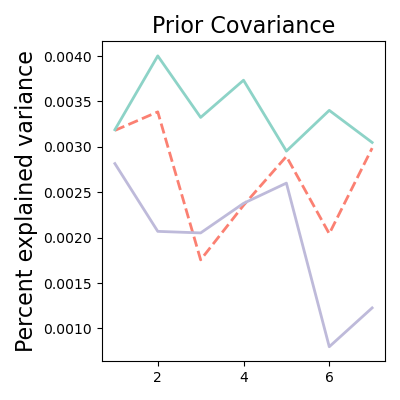}
    \end{subfigure}
    \caption{(\textit{Left}) Estimated direction-specific length scales from running Algorithm \ref{alg:nonstat_val} on the filtered DLPFC data. (\textit{Middle}) Proportion variance explained $pve = \text{diag}(U'SU) / \tr(S)$ with respect to the empirical covariance matrix $S$ comparing loadings matrices $U$ from three models. Note, PCA serves as an upper bound on proportion variance explained of the empirical covariance matrix. (\textit{Right}) Percent variance explained $pve = \text{diag}(U'\Sigma U)/ \tr(\Sigma)$ with respect to the spatial prior covariance matrix $\Sigma$. Please see Appendix Section \ref{sec:appendix_pve} for details.}
    \label{fig:nonstat_DLPFC_filtered}
\end{figure}

The SOFM model was fitted using $k=7$ factors and 400 held-out spots chosen at random on the tissue section. PCA and Laminae-PCA were also fitted using 7 components to provide directly comparable representations. Following Algorithm \ref{alg:nonstat_val}, the domain of sill values was set to $\{1,5,10,50,100,500,1000,5000\}$, and the estimated sill was $\hat{\phi} = 500$ for the filtered and full-gene analyses, and $\hat{\phi} = 100$ for the 3000-HVG analysis. The length scale domain was set to $\{0.5,1,2,4\}$, and the estimates for the filtered analysis are visualized in Figure \ref{fig:nonstat_DLPFC_filtered}. The algorithm ran on 32 cores in $38.51$ seconds for the filtered analysis, $88.92$ seconds for the full-gene analysis, and $18.51$ seconds for the 3000-HVG analysis.

\begin{figure}[htbp]
    \centering
    \begin{subfigure}{\linewidth}
        \centering
        \includegraphics[width=1.\linewidth]{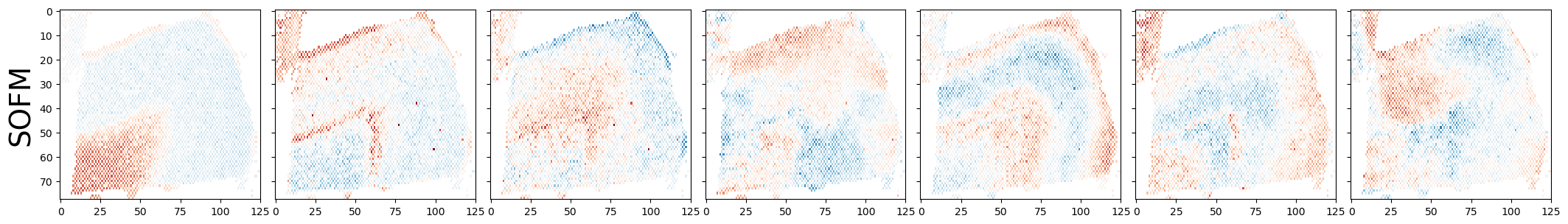}
        \label{fig:loadings_k7}
    \end{subfigure}
    % \begin{subfigure}{\linewidth}
    %     \centering
    %     \includegraphics[width=1.\linewidth]{figs/DLPFC_SOFM_3000HVG_loadings_k7.png}
    % \end{subfigure}
    \begin{subfigure}{\linewidth}
        \centering
        \includegraphics[width=1.\linewidth]{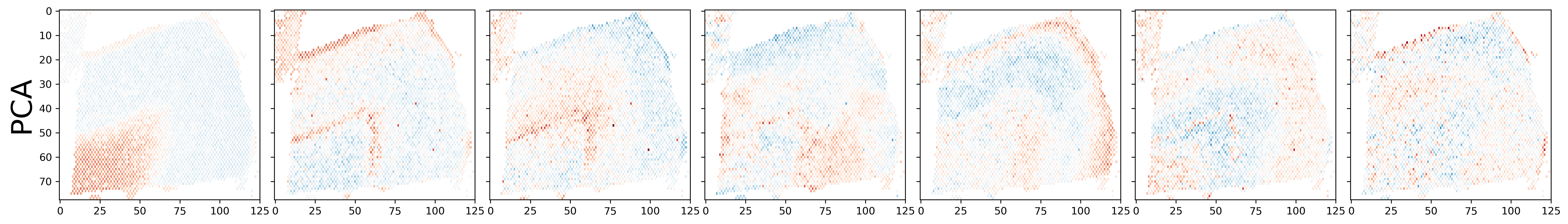}
        \label{fig:pca_loadings}
    \end{subfigure}
    \begin{subfigure}{\linewidth}
        \centering
        \includegraphics[width=1.\linewidth]{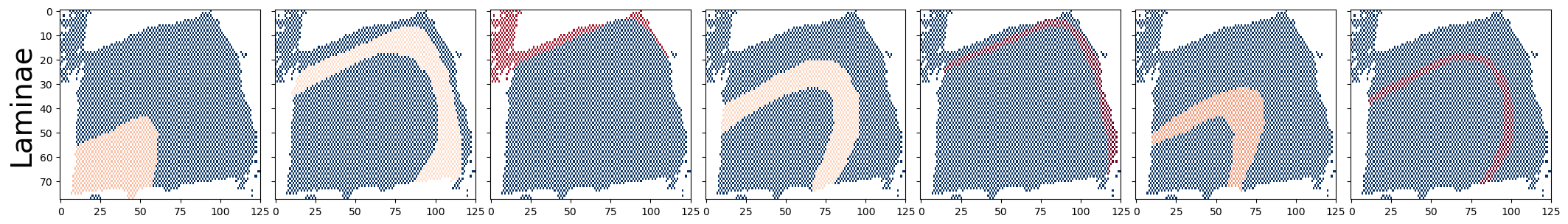}
    \end{subfigure}
    \caption{(Top) SOFM spatial loadings in the DLPFC using the filtered dataset. (Middle) PCA spatial loadings for comparison. (Bottom) Laminae-PCA loadings constructed from the manual cortical-layer and white-matter annotations provided by \citet{maynard2021transcriptome}. The columns show Factors 1--7, ordered by decreasing pseudo-eigenvalue. From left to right, the Laminae-PCA loadings correspond to white matter, layer 3, layer 1, layer 5, layer 2, layer 6, and layer 4. Positive and negative colors indicate opposite factor directions; the sign of each factor is arbitrary.}
    \label{fig:combined_loadings_filtered}
\end{figure}

We report the top 10 genes with the largest positive and negative coefficients for each factor from each of the models in Appendix Section \ref{sec:app_DLPFC}. Using the filtered dataset, the spatial loadings for the three models are shown in Figure \ref{fig:combined_loadings_filtered}. We also report whether \textit{HBB} and \textit{IGKC} occur among the top 10 genes with the largest positive or negative coefficients for the three models in Tables \ref{tab:gene_enrichment} and \ref{tab:gene_enrichment_hvg}, for the filtered and 3000-HVG analyses, respectively. Because the sign of each factor is arbitrary, positive and negative coefficients indicate opposite directions along a factor, but should not be interpreted intrinsically as enrichment or depletion. We report the following exploratory findings:
\begin{itemize}
    \item In Figure \ref{fig:combined_loadings_filtered}, SOFM's first factor shows its largest-magnitude values in the white matter region.  The second factor is associated with both layers 1 and 6, and the third factor is associated with a large portion of layer 5. The fourth factor potentially captures a non-laminar factor described below. The fifth factor contrasts layers 2 and 5, which have positive values (red), with layer 3, which has negative values (blue). The blue to red transition in factor 6 appears to correspond primarily to layer 4.
    \item In Figure \ref{fig:combined_loadings_filtered}, SOFM's Factor 7 appears to be non-laminar and has \textit{HBB} as the gene with the largest negative coefficient (Figure \ref{fig:151675_gene_map_SOFM}). \textit{HBB} was one of the non-laminar spatially variable genes highlighted by \cite{maynard2021transcriptome}.
    \item In the filtered gene analysis, neither \textit{IGKC} nor \textit{HBB} occurs among the 10 genes with the largest positive or negative coefficients for any Laminae-PCA factor (Figure \ref{fig:151675_gene_map_structORACLE}).
    \item SOFM's Factor 6 in the 3000-HVG analysis (Figure \ref{fig:151675_gene_map_SOFM_3000HVGs}) appears to be non-laminar and has \textit{HBB} as the gene with the largest positive coefficient and \textit{IGKC} as the gene with the largest negative coefficient. Thus, \textit{HBB} and \textit{IGKC} contribute in opposite directions to this factor. Four additional immunoglobulin genes also have large negative coefficients on this factor.
    \item SOFM's Factor 7 in the 3000-HVG analysis (Figure \ref{fig:151675_gene_map_SOFM_3000HVGs}) has \textit{HBB} and \textit{IGKC} as the top two genes with the largest negative coefficients.
    \item Figure \ref{fig:nonstat_DLPFC_filtered} shows the spatial distribution of length scales estimated by SOFM. The estimated length scales vary across the tissue and appear qualitatively strongest in the first cortical layer.
    \item Figure \ref{fig:nonstat_DLPFC_filtered} shows that the first three SOFM factors have larger per-factor contributions to the empirical covariance than the corresponding Laminae-PCA factors.
\end{itemize}

While the laminar structure is fundamental for understanding anatomical and functional axes of spatial variation in gene expression in the DLPFC, these results provide evidence of strong additional modes of spatial variation in the data, and are consistent with the non-laminar findings of \cite{maynard2021transcriptome}. These results provide additional evidence that blood- and immune-associated spatial expression patterns may contribute to major modes of covariation alongside patterns associated with neuronal cell populations and anatomical structure. In future research, we plan to develop a formal inferential procedure for identifying genes associated with each SOFM factor, extending beyond the current ranking of genes by their positive and negative coefficients.

\newpage
\begin{table}[htpb]
\centering
\small 
\renewcommand{\arraystretch}{1.3}
\caption{Occurrence of \textit{HBB} and \textit{IGKC} among the top 10 genes with the largest positive or negative coefficients for each factor and method in the filtered gene analysis.}
\label{tab:gene_enrichment}
\begin{tabular}{l | cc | cc | cc}
\hline
 & \multicolumn{2}{c|}{\textbf{SOFM}} & \multicolumn{2}{c|}{\textbf{Laminae}} & \multicolumn{2}{c}{\textbf{PCA}} \\
\cline{2-7} 
\textbf{Gene} & Positive & Negative & Positive & Negative & Positive & Negative \\
\hline
IGKC   & Factor 2 & Factor 4 & & & Factor 2 & \\ \hline
HBB    & & Factor 7 & & & Factor 3& Factor 7 \\ \hline

\end{tabular}
\end{table}
\begin{table}[htpb]
\centering
\small 
\renewcommand{\arraystretch}{1.3}
\caption{Occurrence of \textit{HBB} and \textit{IGKC} among the top 10 genes with the largest positive or negative coefficients for each factor and method in the 3000-HVG analysis.}
\label{tab:gene_enrichment_hvg}
\begin{tabular}{l | cc | cc | cc}
\hline
 & \multicolumn{2}{c|}{\textbf{SOFM (HVG)}} & \multicolumn{2}{c|}{\textbf{Laminae (HVG)}} & \multicolumn{2}{c}{\textbf{PCA (HVG)}} \\
\cline{2-7} 
\textbf{Gene} & Positive & Negative & Positive & Negative & Positive & Negative \\
\hline
IGKC   & \makecell{Factor 2} & \makecell{Factor 3 \\ Factor 6\\ Factor 7} & Factor 6 & & \makecell{Factor 4 \\ Factor 5 \\ Factor 7} & \makecell{Factor 2 \\ Factor 3} \\ \hline
HBB    & \makecell{Factor 5 \\ Factor 6} & \makecell{Factor 4 \\ Factor 7} & Factor 6 & & \makecell{Factor 5 \\ Factor 6 \\ Factor 7} & Factor 4 \\ \hline
\end{tabular}
\end{table}
% \begin{table}[htpb]
% \centering
% \small 
% \renewcommand{\arraystretch}{1.3}
% \caption{Top-10 Gene Enrichment and Depletion Across Models (HVG)}
% \label{tab:gene_enrichment_hvg}
% \begin{tabular}{l | cc | cc | cc}
% \hline
%  & \multicolumn{2}{c|}{\textbf{SOFM (HVG)}} & \multicolumn{2}{c|}{\textbf{PCA (HVG)}} & \multicolumn{2}{c}{\textbf{Laminae (HVG)}} \\
% \cline{2-7} 
% \textbf{Gene} & Enriched & Depleted & Enriched & Depleted & Enriched & Depleted \\
% \hline
% % HLA-A  & & & & & & \\ \hline
% % HLA-B  & & & & & & \\ \hline
% % B2M    & & & & & & \\ \hline
% % CD74   & Factor 7 & \makecell{Factor 2 \\ Factor 6} & & \makecell{Factor 2 \\ Factor 5 \\ Factor 7} & & \\ \hline
% % APP    & & & & & & \\ \hline
% % APOE   & & & & & & \\ \hline
% IGKC   & \makecell{Factor 3 \\ Factor 5} & \makecell{Factor 2 \\ Factor 7} & \makecell{Factor 4 \\ Factor 5 \\ Factor 7} & \makecell{Factor 2 \\ Factor 3} & Factor 6 & \\ \hline
% HBB    & Factor 6 & \makecell{Factor 4 \\ Factor 5 \\ Factor 7} & \makecell{Factor 5 \\ Factor 6 \\ Factor 7} & Factor 4 & Factor 6 & \\ \hline
% % SAA1   & \makecell{Factor 4 \\ Factor 6} & \makecell{Factor 3 \\ Factor 7} & & Factor 6 & \makecell{Factor 2 \\ Factor 5 \\ Factor 7} & \\ \hline
% % IFI27  & Factor 7 & \makecell{Factor 2 \\ Factor 3 \\ Factor 6} & & \makecell{Factor 2 \\ Factor 5 \\ Factor 7} & & \\ \hline
% % HBA2   & & & & & & \\ \hline
% % CLU    & & & & & & \\ \hline
% % CST3   & & & & & & \\ \hline
% % SLC1A2 & & & & & & \\ \hline
% % COX6C  & & & & & & \\ \hline
% % S100B  & & & & & & \\ \hline
% \end{tabular}
% \end{table}

\section{Case Study: NDVI of sub-Saharan conterminous Africa}
\label{sec:Africa}
\subsection{Phenology and the GIMMS data collection}
Phenology is the study of seasonal dynamics and trends of vegetation activity.  In recent years, satellite remote sensing has been used to characterize and monitor dynamics in phenology over both space and time \citep{zhang2003monitoring}.  The normalized difference vegetation index (NDVI) is the most widely used index for studying phenological dynamics from space-borne sensors. This index uses spectral reflectance in two spectral regions that capture photosynthetic activity of plants. The red band (RED; $0.6$-$0.7 \mu m$) corresponds to the spectrum of light absorbed by photosynthetically active biomass, whereas, the near-infrared radiation (NIR; $0.7$-$1.1$ $\mu m$) is strongly reflected by healthy photosynthetic biomass. NDVI is calculated as follows,
\[
\mathrm{NDVI} = \frac{\mathrm{NIR}-\mathrm{RED}}{\mathrm{NIR}+\mathrm{RED}}
\]

The Advanced Very-High-Resolution Radiometer (AVHRR) has been collecting images in these spectral bands across multiple satellite missions since 1981 \citep{ehrlich1994applications}. These  images are preprocessed and productized into a collection of bi-weekly NDVI global measurements at $1/12$ degrees spatial resolution known as the Global Inventory Modeling and Mapping Studies (GIMMS) \citep{pinzon2023global}. This dataset is particularly attractive because the NDVI bi-weekly measurements are based on a composite of images collected daily within each bi-weekly interval. As a result, pixel values in images corrupted by atmospheric effects, particularly clouds, are systematically removed yielding high fidelity decadal-scale time series of terrestrial ecosystem dynamics \citep{fensholt2012evaluation}.

% Sub-Saharan Africa is an important study region for a number of reasons. The first is that the region traverses large-scale climate regions: arid, semi-arid, humid tropics, and sub-humid tropics. As such, the subcontinent is important for understanding how global warming is affecting each of these climate-regions \citep{keenan2014net}. The second reason is many regions of the continent are agriculturally constrained which has implications for feeding populations as well as for resulting conflict \citep{bolton2013forecasting,hsiang2013quantifying}.
\subsection{Scientific question of interest}
\textbf{Here is our scientific question of interest: by conditioning on long-range spatial dependence of phenological signatures at continental scale, can SOFM estimate useful spatial modes of temporal variation on short time scales (e.g. 1 year)?} This use case is important for scientists who wish to understand large-scale environmental patterns and dynamics when data may only be available for a short time period. 

PCA may struggle in these settings due to the high spatial dimension (the 1-year study has $239318$ spatial locations and $24$ temporal observations), whereas the regularization imposed by the spatial prior in SOFM may prove useful. For this analysis we examine dynamics in NDVI at continental scale over Africa.  We use the full 40-year time series as a benchmark to compare our model results from fitting to a single year of data. The 40-year data will help give a sense of major climate regions in general, whereas results from the last year in the dataset may have departures due to noise variation in the climate system, due to climate change (that is, change in the trends of the climate system) \citep{piao2019plant}. To that end, we run SOFM and PCA on a single year of GIMMS data and compare the results to running PCA on the full 40 years of data ($984$ temporal observations). Specifically, we compare results from 01/08/2022-12/24/2022 in the single year study, and from 01/08/1982-12/24/2022 in the 40 year study. Both studies model sub-Saharan conterminous Africa, and the data are processed by subtracting the mean spatial vector from each time point, as assumed under both the SOFM and PCA models.

\subsection{Spatially orthogonal phenological regions in sub-Saharan Africa}
\begin{figure}
    \centering
    \begin{subfigure}{0.475\linewidth}
        \centering
        \includegraphics[width=\linewidth]{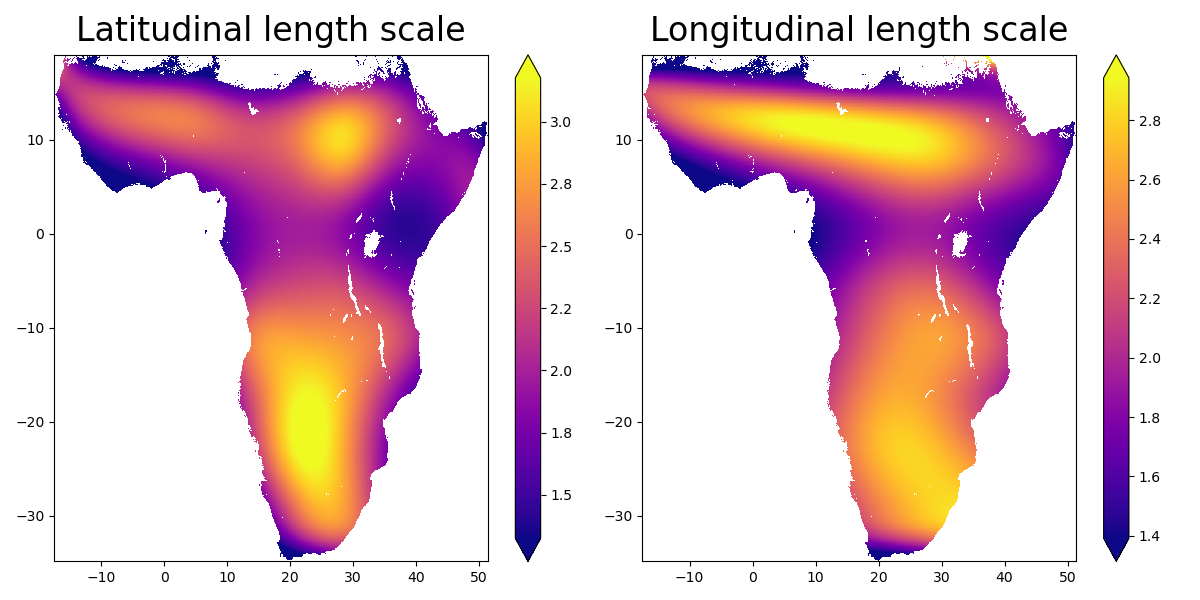}
    \end{subfigure}
    \begin{subfigure}{0.25\linewidth}
        \centering
        \includegraphics[width=\linewidth]{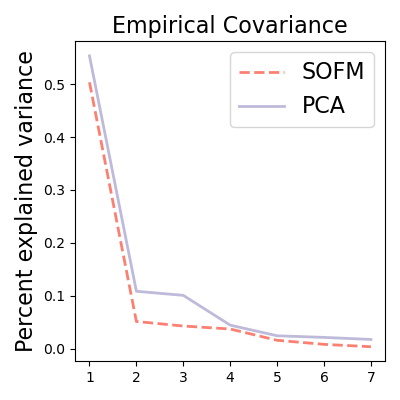}
    \end{subfigure}
    \begin{subfigure}{0.25\linewidth}
        \centering
        \includegraphics[width=\linewidth]{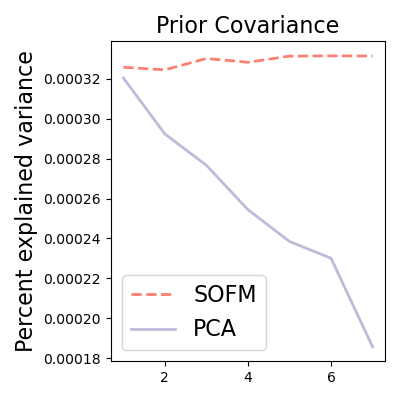}
    \end{subfigure}
    \caption{(Left) Latitudinal and longitudinal length scales estimated by SOFM. Dependence ranges from 0 to 3 degrees throughout the subcontinent. (Right) Percent variance explained plots for the empirical covariance matrix and the spatial prior covariance matrix.}
    \label{fig:africa_lat_lon_depend_eig}
    \end{figure}

%Discuss stronger length scales in the Sahel region, weaker in transition regions
The estimates of SOFM's nonstationary spatial covariance are presented in Figure \ref{fig:africa_lat_lon_depend_eig}. In the left side of the figure, latitudinal length scales $\lambda_{lat,i}$ are plotted for each $i$th pixel and the longitudinal length scales $\lambda_{lon,i}$ are plotted on the right.  In both directions of dependence, we see stronger spatial dependence in the Sahel and sub-humid tropics, whereas climate transition regions from the Sahara to the Sahel and from the Sahel to the humid tropics, where climate gradients are weaker, there is less spatial dependence.

\begin{figure}[htbp]
    \centering
    \begin{subfigure}{\linewidth}
        \centering
        \includegraphics[width=1.\linewidth]{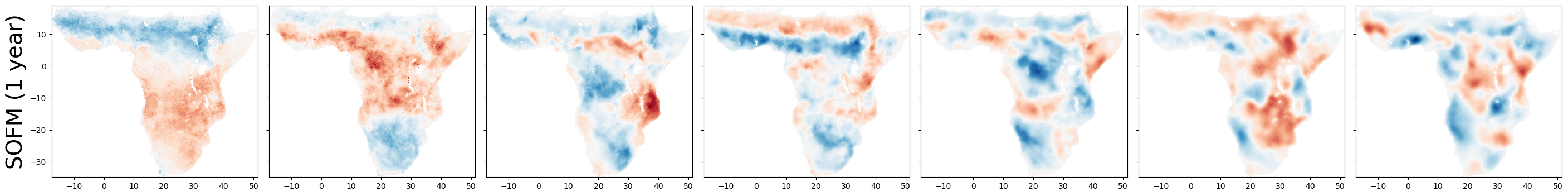}
        \label{fig:africa_loadings_k7}
    \end{subfigure}
    \begin{subfigure}{\linewidth}
        \centering
        \includegraphics[width=1.\linewidth]{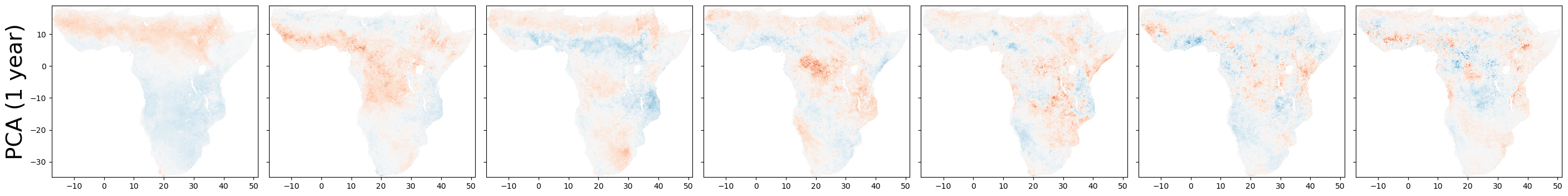}
        \label{fig:africa_pca_loadings}
    \end{subfigure}
    \begin{subfigure}{\linewidth}
        \centering
        \includegraphics[width=1.\linewidth]{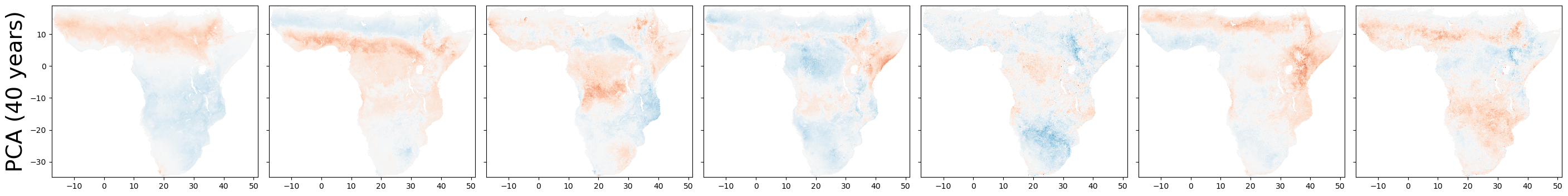}
        \label{fig:africa_pca_loadings_full}
    \end{subfigure}
    \caption{(Top) SOFM run on a single year of data. (Middle) PCA run on a single year of data. (Bottom) PCA run on 40 years of data.}
    \label{fig:africa_loadings_comparison}
\end{figure}

%Discuss percent explained variance for SOFM is just below PCA, but maintains spatial variance as number of factors increases whereas PCA dies off.
Continuing our analysis of Figure \ref{fig:africa_lat_lon_depend_eig}, the variance explained by the SOFM model follows closely that of PCA because the PCA represents the upper bound on this metric. When we inspect the amount of variance explained \textit{with respect to the spatial prior}, the SOFM model is relatively stable, whereas PCA components decrease to $0$. The first component of both models captures about 50 percent of the total variance. Cross-checking Figure \ref{fig:africa_loadings_comparison}, the first factor corresponds to anti-correlated seasonal dynamics of the northern and southern hemispheres of the continent.  The proportion of explained variance for the remaining factors drops off quickly. However, many of those factors are spatially similar to the factors found in the model estimated from the full 40-year time series. 

%Explain that the humid forest transition region is split by SOFM, but not found by PCA, even though it is a well known region confirmed by 40 year data.
By comparing results from the model estimated from a single year of data with results from the model estimated using 40-years of data, we find that SOFM's single year model identifies a novel spatial pattern not present in corresponding results from PCA.  In Figure \ref{fig:africa_loadings_comparison}, the second factor from the 40-year PCA model shows loadings that have the largest magnitudes in the humid-tropics. However, the corresponding results from the 1-year SOFM model include two subregions identified in the 40-year PCA model that are not captured by the 1-year PCA model: a subregion in the eastern humid tropics near South Sudan and another in Mozambique. While the 1-year PCA model does not capture these regions from the 40-year PCA model, SOFM does capture these missing regions in its third factor, demonstrating that spatial regularization allows SOFM to find underlying spatial factors that PCA does not find, and those factors are corroborated by long term 40-year patterns. An interesting next step in this analysis would be to understand why SOFM breaks the humid tropics into these two regions compared to the 40 year model. We leave this for future research.

% \begin{figure}
%     \centering
%     \includegraphics[width=0.5\linewidth]{figs/Africa_timeseries_k7.png}
%     \caption{The posterior expectation of the latent factors $\E[z | y,U^{\dagger},L^{\dagger},\sigma^{2\dagger}]$ from the SOFM model.}
%     \label{fig:africa_timeseries}
% \end{figure}
\section{Discussion}
\label{sec:discussion}
In this work, we derived the sampling distribution of the singular value decomposition of a spatially-distributed multivariate matrix under the repeated eigenvalue assumption in the principal component framework (Proposition \ref{prop:prior}).  In applied settings, this model is useful as the loadings themselves tend to be spatially distributed. Using this sampling distribution as a prior on orthogonal loadings of a probabilistic PCA model, we prove the maximum a posteriori orthogonal loadings are the eigenvectors of  
$(S + \frac{1}{n} \Sigma)$ (Theorem \ref{thm:MAP_U}). This theorem is useful for interpretative purposes, as well as for methodological purposes as estimation can be carried out once for a larger number of factors and then pruned as necessary.  As the spatial dimension of our applications is large and the eigendecomposition of $(S + \frac{1}{n} \Sigma)$ is not feasible, we provide a probabilistic PCA approach to inference and estimation, yielding a linear-time algorithm with respect to the number of spatial locations for MAP estimation of the orthogonal loadings.

The problem of estimating the underlying spatial dependence structure is also addressed in this work.  Under smoothness assumptions of the latitudinal and longitudinal length scales of a nonstationary covariance model, we extend our methodology for MAP inference of the orthogonal loadings when spatial locations are held-out.  Using this approach, the held-out predictive distribution is maximized with respect to the length scale parameters. To our knowledge, this approach for estimating the underlying nonstationary spatial covariance is the first of its kind in the application areas covered in this paper.

Our resulting spatially orthogonal factor model provided interpretable results in both case studies.  In the spatial transcriptomics case study in the dorsolateral prefrontal cortex brain region, our model identified spatial factors associated with cortical layers, as well as non-laminar blood- and immune-associated gene expression patterns, consistent with findings from \cite{maynard2021transcriptome}. The corresponding gene coefficients also identified additional genes associated with these factors.  In the environmental science case study, we ran SOFM on a single year of NDVI data to demonstrate the benefits of prior spatial estimation as well as spatial regularization in this setting. We checked these results by showing the resulting phenological regions were not inconsistent with a related model fit to 40 years of data.

Moving forward, in the environmental science applications, there are opportunities to extend our model to incorporate temporal dependence in the latent variable or to derive a new prior distribution where the loadings themselves evolve through time. In the spatial transcriptomics setting, there is an opportunity to extend our methodology by developing formal inferential procedures for identifying genes associated with each SOFM factor.
\section{Code availability}
Code for running SOFM can be found at \hyperlink{https://github.com/daniel-s-cunha/SOFM}{https://github.com/daniel-s-cunha/SOFM} and the software package can be installed using \texttt{pip install sofm}.

\bibliography{Refs.bib}
\bibliographystyle{apalike}
\newpage
\section{Appendix: DLPFC results}
\label{sec:app_DLPFC}
\begin{figure}
    \centering
    \includegraphics[width=0.45\linewidth]{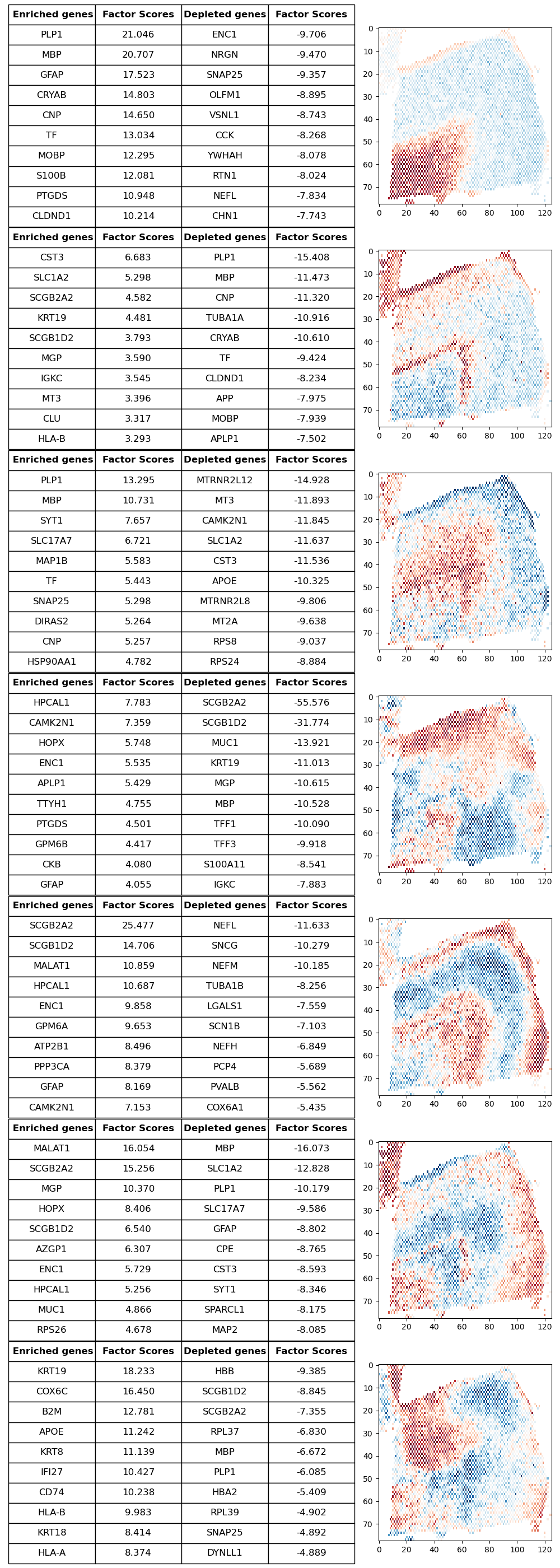}
    \caption{Genes with the largest positive and negative coefficients for the SOFM model.}
    \label{fig:151675_gene_map_SOFM}
\end{figure}

\begin{figure}
    \centering
    \includegraphics[width=0.45\linewidth]{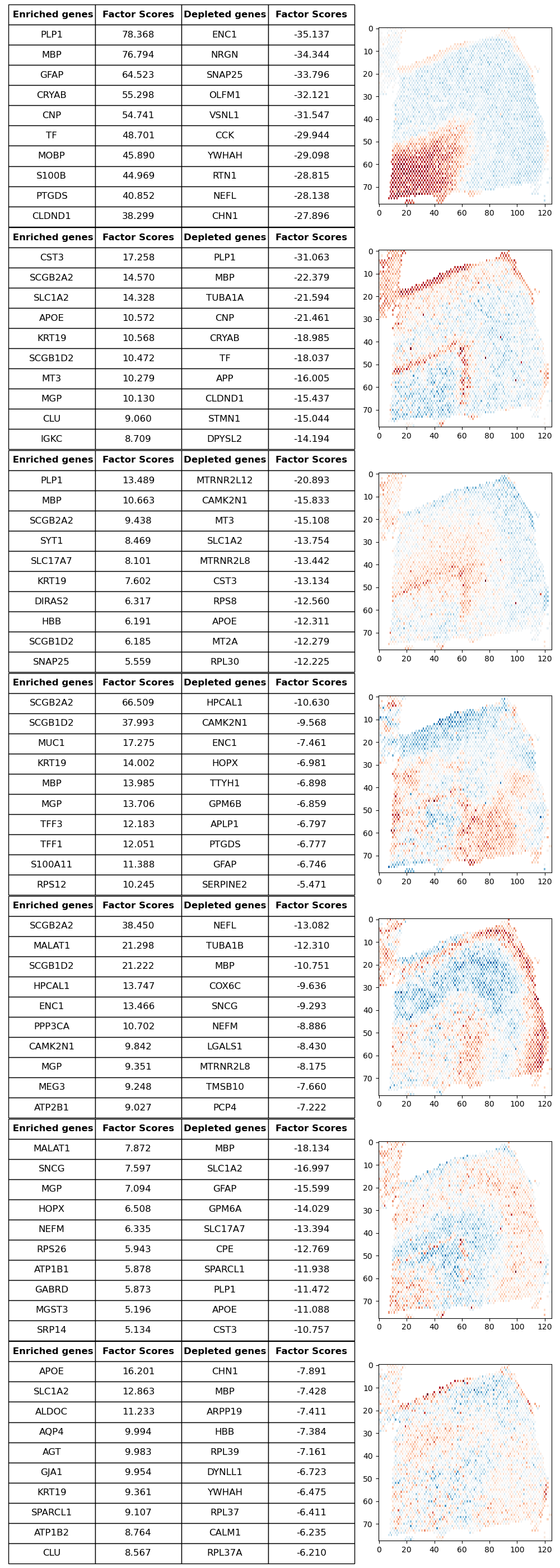}
    \caption{Genes with the largest positive and negative coefficients for regular PCA.}
    \label{fig:151675_gene_map_regPCA}
\end{figure}

\begin{figure}
    \centering
    \includegraphics[width=0.45\linewidth]{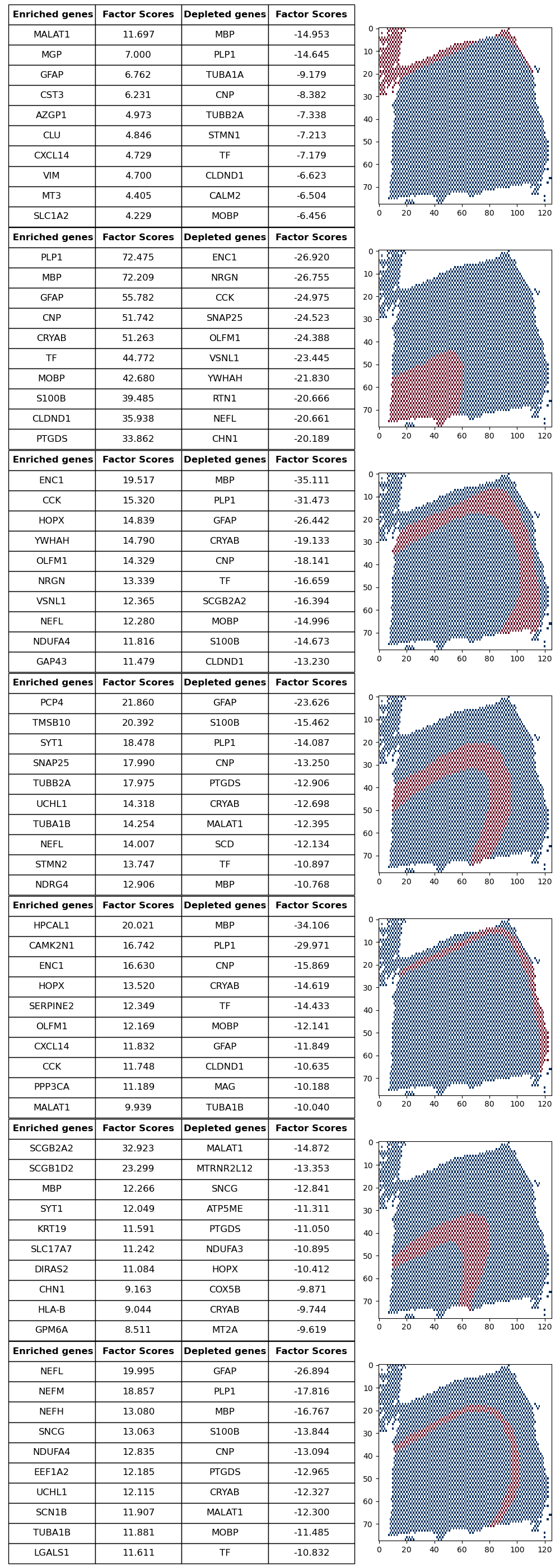}
    \caption{Genes with the largest positive and negative coefficients for Laminae-PCA.}
    \label{fig:151675_gene_map_structORACLE}
\end{figure}

\begin{figure}
    \centering
    \includegraphics[width=0.45\linewidth]{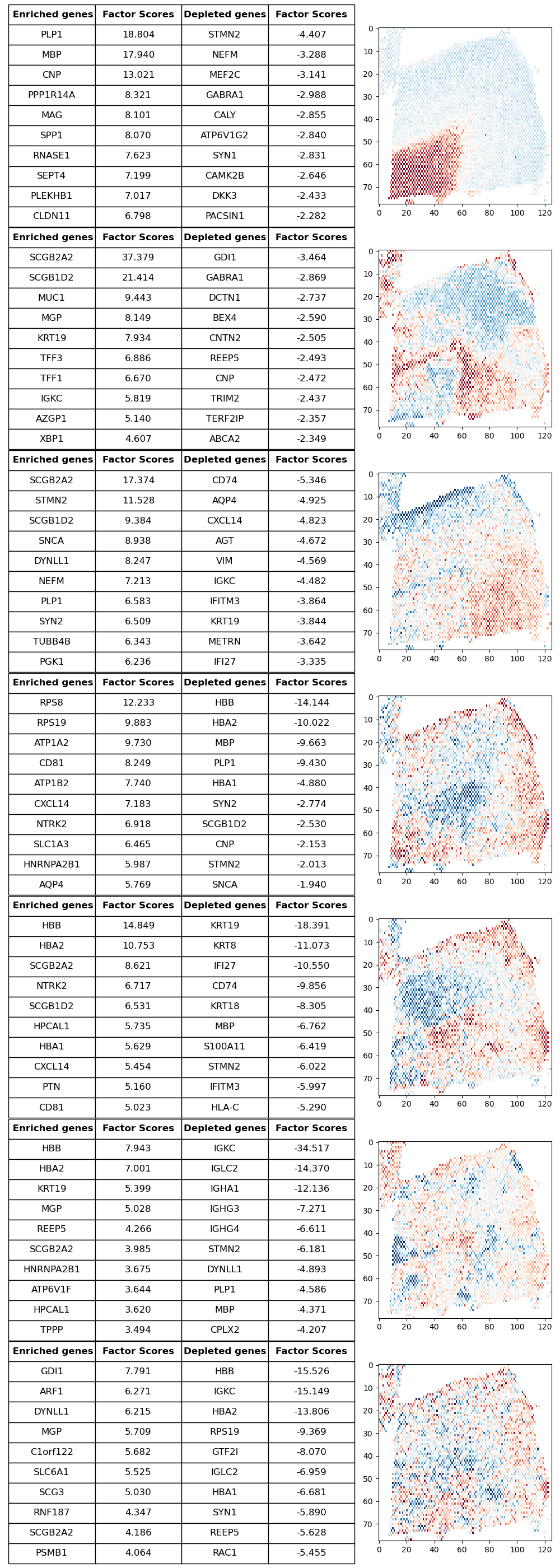}
    \caption{Genes with the largest positive and negative coefficients for the  SOFM model fitted using the top 3000 highly variable genes (HVGs).}
    \label{fig:151675_gene_map_SOFM_3000HVGs}
\end{figure}

\begin{figure}
    \centering
    \begin{subfigure}{0.475\linewidth}
        \centering
        \includegraphics[width=\linewidth]{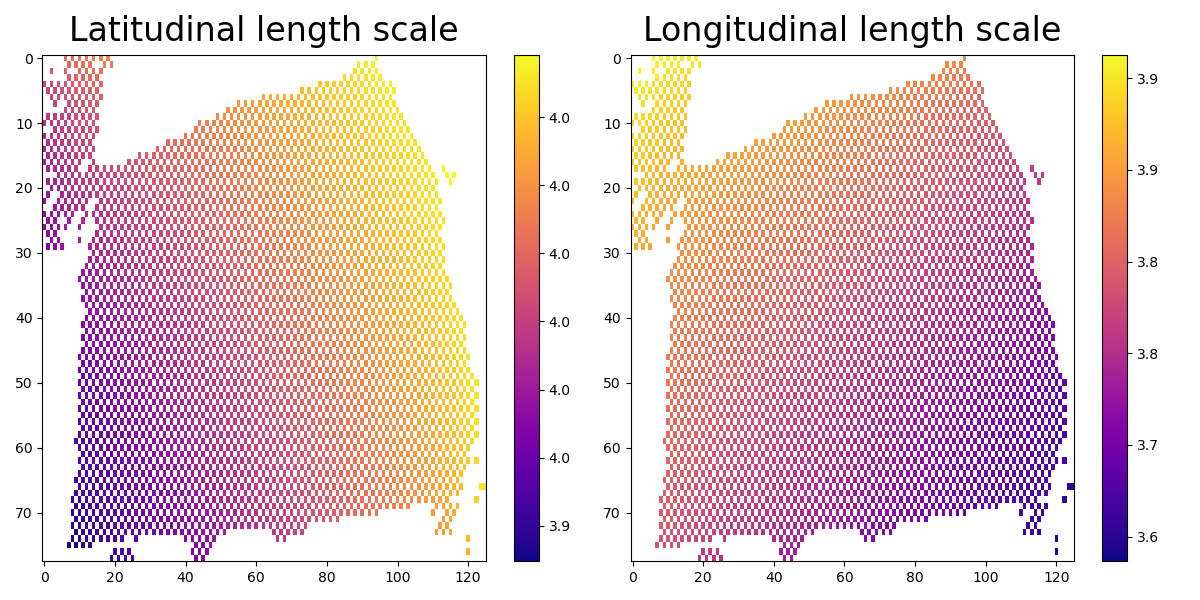}    
    \end{subfigure}
    \begin{subfigure}{0.25\linewidth}
        \centering
        \includegraphics[width=\linewidth]{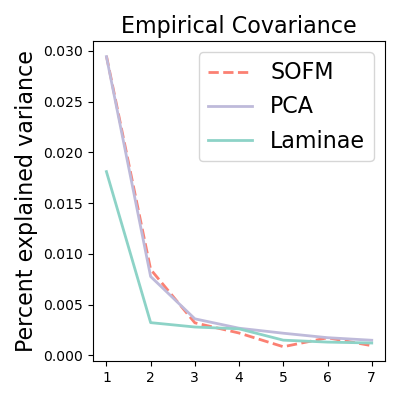}
    \end{subfigure}
    \begin{subfigure}{0.25\linewidth}
        \centering
        \includegraphics[width=\linewidth]{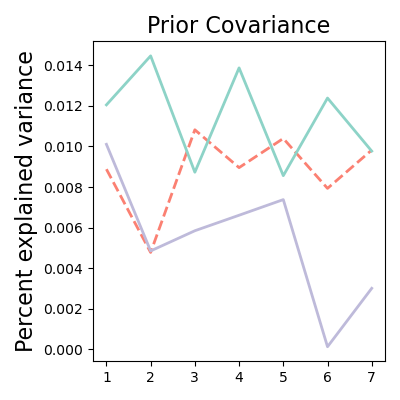}
    \end{subfigure}
    \caption{(\textit{Left}) Estimated direction-specific length scales from running Algorithm \ref{alg:nonstat_val} using the full-gene (33,538 genes) DLPFC dataset. (\textit{Middle}) Proportion variance explained $pve = \text{diag}(U'SU) / \tr(S)$ with respect to the empirical covariance matrix $S$ comparing loadings matrices $U$ from three models. Note, PCA serves as an upper bound on proportion variance explained of the empirical covariance matrix. (\textit{Right}) Percent variance explained $pve = \text{diag}(U'\Sigma U)/ \tr(\Sigma)$ with respect to the spatial prior covariance matrix $\Sigma$. Please see Appendix Section \ref{sec:appendix_pve} for details.}
    \label{fig:nonstat_DLPFC_full_gene}
\end{figure}

\begin{figure}[htbp]
    \centering
    \begin{subfigure}{\linewidth}
        \centering
        \includegraphics[width=1.\linewidth]{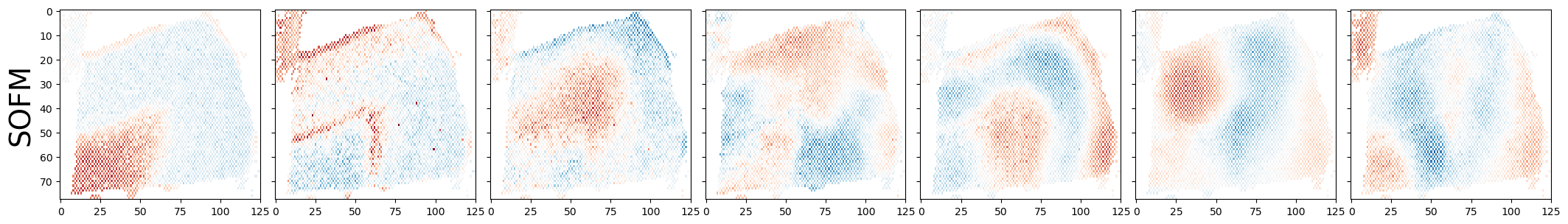}
        \label{fig:loadings_k7}
    \end{subfigure}
    % \begin{subfigure}{\linewidth}
    %     \centering
    %     \includegraphics[width=1.\linewidth]{figs/DLPFC_SOFM_3000HVG_loadings_k7.png}
    % \end{subfigure}
    \begin{subfigure}{\linewidth}
        \centering
        \includegraphics[width=1.\linewidth]{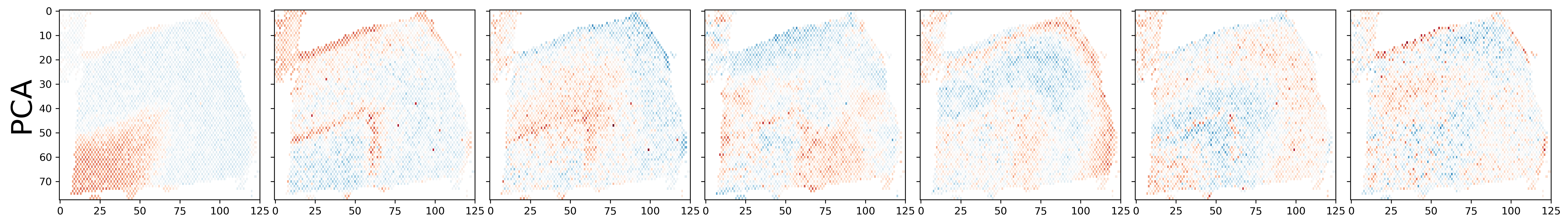}
        \label{fig:pca_loadings}
    \end{subfigure}
    \begin{subfigure}{\linewidth}
        \centering
        \includegraphics[width=1.\linewidth]{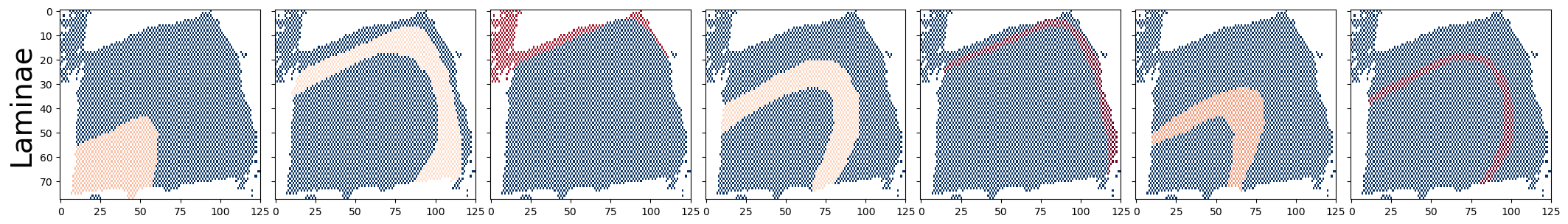}
    \end{subfigure}
    \caption{Results using the full-gene (33,538 genes) DLPFC dataset. (Top) SOFM spatial loadings in the DLPFC. (Middle) PCA spatial loadings for comparison. (Bottom) Laminae-PCA loadings constructed from the manual cortical-layer and white-matter annotations provided by \citet{maynard2021transcriptome}. The columns show Factors 1--7, ordered by decreasing pseudo-eigenvalue. From left to right, the Laminae-PCA loadings correspond to white matter, layer 3, layer 1, layer 5, layer 2, layer 6, and layer 4. Positive and negative colors indicate opposite factor directions; the sign of each factor is arbitrary.}
    \label{fig:combined_loadings_full_gene}
\end{figure}

Figures \ref{fig:151675_gene_map_SOFM}, \ref{fig:151675_gene_map_regPCA}, \ref{fig:151675_gene_map_structORACLE}, and \ref{fig:151675_gene_map_SOFM_3000HVGs} provide a summary of the spatial loadings and genes with the largest positive and negative coefficients for SOFM, regular PCA, and Laminae-PCA on the filtered gene set, as well as SOFM using the top 3000 HVGs. Note that since the sign of each factor is arbitrary, positive and negative coefficients indicate opposite directions along a factor, but should not be interpreted intrinsically as enrichment or depletion. In addition, Figures \ref{fig:nonstat_DLPFC_full_gene} and \ref{fig:combined_loadings_full_gene} provide results for the full-gene analysis (33,538 genes), which may be compared to the main results shown in Figures \ref{fig:nonstat_DLPFC_filtered} and \ref{fig:combined_loadings_filtered} for the filtered analysis (14,638 genes).

\clearpage

\section{Appendix: Proofs of Proposition \ref{prop:prior} and Theorem \ref{thm:MAP_U}}
This section contains the proof for Proposition \ref{prop:prior}.
\subsection{Derivation of Jacobian for prior distribution}
The derivation is broken into three successive Jacobian steps.  The first is the SVD transformation of an $\nu \times m$ matrix, the second is inverting the singular values, and the third is parameterizing the singular values to align with the PCA model formulation. 

To that end, let $Y^{(0)} = V D U'$ where $V\in O(\nu)$, $U \in O(m)$, and the upper $m\times m$ entries of $D$ are diagonal, and the lower $(\nu-m)\times m$ entries are zero. That is, we handle the case $\nu \geq m$ here. Take the differential of $Y^{(0)}$,
\[
dY^{(0)} = dV D U' + V dD U' + V D dU'
\]
Left multiply by $V'$ and right multiply by $U$ to isolate the diagonal entries,
\[
V'dY^{(0)} U = (V'dV) D + dD + D (U'dU)'
\]
Since $V$ is orthogonal we have $V'V = I_{\nu}$, and thus $dV' V + V' dV = 0$, yielding the skew symmetric relationship $V'dV = -(V'dV)'$. The same follows for $U$. Taking the exterior product $(V'dY^{(0)} U) = \det(V)^{m} \det(U)^{\nu} (dY^{(0)}) = (dY^{(0)})$ since $U,V$ are orthogonal. Define $\Omega = V'dV$ and $\Psi = U'dU$ for notational simplicity.  We need to isolate all degrees of freedom in $V' dY^{(0)} U$ and take the exterior product over those degrees of freedom.  To that end, break $V$ and $U$ into three and two sets of columns $V = (V_{1:k},V_{k+1:m},V_{m+1:\nu})$ and $U = (U_{1:k},U_{k+1:m})$.  Then, 
\begin{align*}
    V' dY^{(0)} U = \begin{bmatrix}
        V_{1:k}' dY^{(0)} U_{1:k} & V_{1:k}' dY^{(0)} U_{k+1:m} \\
        V_{k+1:m}' dY^{(0)} U_{1:k} & V_{k+1:m}' dY^{(0)} U_{k+1:m} \\
        V_{m+1:\nu}' dY^{(0)} U_{1:k} & V_{m+1:\nu}' dY^{(0)} U_{k+1:m}
    \end{bmatrix}
\end{align*}
We will evaluate the exterior product of each block separately.
\subsubsection{Exterior product of top left block}
The top left block $V_{1:k}' dY^{(0)} U_{1:k}$ has full degrees of freedom. On the diagonal, we simply have $\bigwedge_{i = 1}^k dD_{ii}$. On the off-diagonal, we have by skew-symmetry,
\begin{align*}
    \bigwedge_{i\ne j}(V_{1:k}' dY^{(0)} U_{1:k}) &= \left(\bigwedge_{i = 1}^k \bigwedge_{j=i+1}^k (\Omega_{ij}D_{jj} - D_{ii}\Psi_{ij})\right)\left(\bigwedge_{i = 1}^k \bigwedge_{j=1}^{i-1} (\Omega_{ij}D_{jj} - D_{ii}\Psi_{ij})\right)\\
    &= \bigwedge_{i = 1}^k \bigwedge_{j=i+1}^k (\Omega_{ij}D_{jj} - D_{ii}\Psi_{ij}) (-\Omega_{ij}D_{ii} + D_{jj}\Psi_{ij})\\
    &= \bigwedge_{i = 1}^k \bigwedge_{j=i+1}^k (\Omega_{ij}\Psi_{ij}D_{jj}^2 - \Omega_{ij}\Psi_{ij}D_{ii}^2)\\
    &= \prod_{j>i}^k \left(D_{jj}^2-D_{ii}^2\right) \bigwedge_{i = 1}^k \bigwedge_{j=i+1}^k \Omega_{ij}\Psi_{ij}
\end{align*}
\subsubsection{Exterior product of middle left and top right blocks}
Now we take the exterior product of the middle left and top right blocks, which do not have any diagonal entries. 
\begin{align*}
    \bigwedge(V_{k+1:m}' dY^{(0)} U_{1:k})*\\
    \bigwedge(V_{1:k}' dY^{(0)} U_{k+1:m}) &= \bigwedge_{i = k+1}^m \bigwedge_{j=1}^k (\Omega_{ij}D_{jj} - D_{k+1,k+1}\Psi_{ij})\bigwedge_{i = 1}^k \bigwedge_{j=k+1}^{m} (\Omega_{ij}D_{k+1,k+1} - D_{ii}\Psi_{ij})\\
    &= \bigwedge_{i = k+1}^m \bigwedge_{j=1}^k (\Omega_{ij}D_{jj} - D_{k+1,k+1}\Psi_{ij})(-\Omega_{ij}D_{k+1,k+1} + D_{jj}\Psi_{ij})\\
    &= \bigwedge_{i = k+1}^m \bigwedge_{j=1}^k \Omega_{ij}\Psi_{ij} D_{jj}^2 - \Omega_{ij}\Psi_{ij} D_{k+1,k+1}^2\\
    &= \prod_{j=1}^{k} (D_{jj}^2 - D_{k+1,k+1}^2)^{(m-k)} \bigwedge_{i = k+1}^m \bigwedge_{j=1}^k \Omega_{ij}\Psi_{ij}
\end{align*}
\subsubsection{Exterior product of middle right block}
The middle right block has both diagonal and off-diagonal degrees of freedom for consideration. However, since the diagonal term is the same for all subdiagonal and superdiagonal entries of this block, we only consider the subdiagonal degrees of freedom by skew-symmetry of $\Omega$ and $\Psi$.
\begin{align*}
    \bigwedge V_{k+1:m}' dY^{(0)} U_{k+1:m} &= (dD_{k+1,k+1}) \bigwedge_{i = k+1}^m \bigwedge_{j=i+1}^m D_{k+1,k+1}(\Omega_{ij} - \Psi_{ij})\\
    &= (dD_{k+1,k+1}) D_{k+1,k+1}^{\binom{m-k}{2}}\bigwedge_{i = k+1}^m \bigwedge_{j=i+1}^m (\Omega_{ij} - \Psi_{ij})
\end{align*}
\subsubsection{Exterior product of the bottom left block}
\begin{align*}
    \bigwedge (V_{m+1:\nu}' dY^{(0)} U_{1:k}) &= \bigwedge_{i = m+1}^\nu \bigwedge_{j=1}^k (\Omega_{ij}D_{jj} - 0 \Psi_{ij})\\
    &= \prod_{j = 1}^k D_{jj}^{\nu-m} \bigwedge_{i = m+1}^\nu \bigwedge_{j=1}^k \Omega_{ij}
\end{align*}

\subsubsection{Exterior product of the bottom right block}
\begin{align*}
    \bigwedge (V_{m+1:\nu}' dY^{(0)} U_{k+1:m}) &= \bigwedge_{i = m+1}^\nu \bigwedge_{j=k+1}^m (\Omega_{ij}D_{k+1,k+1} - 0 \Psi_{ij})\\
    &= D_{k+1,k+1}^{(\nu-m)(m-k)} \bigwedge_{i = m+1}^\nu \bigwedge_{j=k+1}^m \Omega_{ij}
\end{align*}
\subsubsection{Wedging the blocks together}
Wedging each of the 6 blocks together yields,
\begin{align*}
     (dY^{(0)}) &= \prod_{j>i}^k \left(D_{jj}^2-D_{ii}^2\right)\prod_{j=1}^{k} (D_{jj}^2 - D_{k+1,k+1}^2)^{(m-k)} D_{k+1,k+1}^{(\nu-m)(m-k)+\binom{m-k}{2}} \prod_{j = 1}^k D_{jj}^{\nu-m}\\
     &\quad (dD) \bigwedge_{i = m+1}^\nu \bigwedge_{j=1}^m \Omega_{ij} \bigwedge_{i = k+1}^m \bigwedge_{j=i+1}^m (\Omega_{ij} - \Psi_{ij}) \bigwedge_{i = k+1}^m \bigwedge_{j=1}^k \Omega_{ij}\Psi_{ij}\bigwedge_{i = 1}^k \bigwedge_{j=i+1}^k \Omega_{ij}\Psi_{ij}
\end{align*}
\subsubsection{Jacobian of inverse transformation}
The next step in prior distribution development is to invert the upper $m\times m$ block diagonal of $D$, which has $k+1$ degrees of freedom. Let $B_{jj} = D_{jj}^{-1}$ for $j = 1,\dots,k+1$ and $0$ otherwise. For each of those $k+1$ transformations, we have the Jacobian $dD_{jj} = -B_{jj}^{-2} dB_{jj}$.  Thus, taking the exterior product of both sides results in, 
\begin{align*}
    (dD) &= \bigwedge_{j=1}^{k+1} dD_{jj} \\
    &= \bigwedge_{j=1}^{k+1} (-B_{jj}^{-2} dB_{jj})\\
    &= \left(\prod_{j=1}^{k+1} B_{jj}^{-2}\right) (dB)
\end{align*}
Plugging this back into the full Jacobian transformation, we have,
\begin{align*}
     (dY^{(0)}) &= \prod_{j>i}^k \left(B_{jj}^{-2}-B_{ii}^{-2}\right)\prod_{j=1}^{k} (B_{jj}^{-2} - B_{k+1,k+1}^{-2})^{(m-k)} B_{k+1,k+1}^{-(\nu-m)(m-k)-\binom{m-k}{2}-2} \prod_{j = 1}^k B_{jj}^{-(\nu-m)-2}\\
     &\quad (dB)\bigwedge_{i = m+1}^\nu \bigwedge_{j=1}^m \Omega_{ij} \bigwedge_{i = k+1}^m \bigwedge_{j=i+1}^m (\Omega_{ij} - \Psi_{ij}) \bigwedge_{i = k+1}^m \bigwedge_{j=1}^k \Omega_{ij}\Psi_{ij}\bigwedge_{i = 1}^k \bigwedge_{j=i+1}^k \Omega_{ij}\Psi_{ij}
\end{align*}
Note that after this part of the transformation, we think of $UB^{-2}U'$ as a precision matrix as opposed to a covariance matrix and as such we will also swap notation for the corresponding precision parameter, writing $\Sigma$ in place of $\Phi^{-1}$ to denote its new interpretation.
\subsubsection{Jacobian of PCA transformation}
The last transformation is needed to create alignment with the PCA model formulation.  To that end, define the $k\times k$ diagonal matrix $L$ as a mapping of the first $k$ singular values and the parameter $\sigma$ as a mapping of the $k+1$ singular value as follows,
\[
B_{jj} = \begin{cases}
    (L_{jj}^2 + \sigma^2)^{\frac{1}{2}}, \text{for $j = 1,\dots,k$} \\
    (\sigma^2)^{\frac{1}{2}}, \text{for $j = k+1$} 
\end{cases}
\]
With a Jacobian for $j = 1,...,k$
\[
dB_{jj} = d(L_{jj}^2 + \sigma^2)^{\frac{1}{2}} = \frac{1}{2}(L_{jj}^2 + \sigma^2)^{-\frac{1}{2}}(2 L_{jj} dL_{jj} + d\sigma^2)
\]
And for $k+1$,
\[
dB_{k+1,k+1} = d (\sigma^2)^{\frac{1}{2}} = \frac{1}{2} (\sigma^2)^{-\frac{1}{2}} d\sigma^2
\]
%First derive (dB) and show how the d\sigma^2 cancel except for (dL)(d\sigma^2)
In order to plug these results back into the final equation, there are a few steps.  First note, 
\begin{align*}
    (dB) &= \bigwedge_{j = 1}^{k+1} dB_{jj}\\
    &= \bigwedge_{j = 1}^{k} \frac{1}{2}(L_{jj}^2 + \sigma^2)^{-\frac{1}{2}}(2 L_{jj} dL_{jj} + d\sigma^2) \wedge \frac{1}{2} (\sigma^2)^{-\frac{1}{2}} d\sigma^2
\end{align*}
Only the product of the $dL$ terms and a single $d\sigma^2$ remain since $d\sigma^2 \wedge d\sigma^2 = 0$, thus,
\begin{align*}
    (dB) &= \bigwedge_{j = 1}^{k} (L_{jj}^2 + \sigma^2)^{-\frac{1}{2}}(L_{jj} dL_{jj}) \wedge \frac{1}{2} (\sigma^2)^{-\frac{1}{2}} d\sigma^2\\
    &= \left(\prod_{j = 1}^{k}(L_{jj}^2 + \sigma^2)^{-\frac{1}{2}}L_{jj}\right) \frac{1}{2}(\sigma^2)^{-\frac{1}{2}} (dL)(d\sigma^2)
\end{align*}
%Check the above is correct, since it's late...
%Then show how each of the main terms in the Jacobian is transformed
There are a number of terms to evaluate for the transformation,
\begin{align*}
\prod_{j>i}^k \left(B_{jj}^{-2}-B_{ii}^{-2}\right) &= \prod_{j>i}^k \left((L_{jj}^2 + \sigma^2)^{-1} - (L_{ii}^2 + \sigma^2)^{-1}\right)
\end{align*}
The next term,
\begin{align*}
    \prod_{j=1}^{k} (B_{jj}^{-2} - B_{k+1,k+1}^{-2})^{(m-k)} &= \prod_{j=1}^{k} ((L_{jj}^2 + \sigma^2)^{-1} - \sigma^{-2})^{(m-k)}
\end{align*}
The next term,
\begin{align*}
    B_{k+1,k+1}^{-(\nu-m)(m-k)-\binom{m-k}{2}-2} &= (\sigma^{2*\frac{1}{2}})^{-(\nu-m)(m-k)-\binom{m-k}{2}-2}
\end{align*}
The last term,
\begin{align*}
    \prod_{j = 1}^k B_{jj}^{-(\nu-m)-2} &= \prod_{j = 1}^k ((L_{jj}^2 + \sigma^2)^{\frac{1}{2}})^{-(\nu-m)-2}\\
    &= \prod_{j = 1}^k (L_{jj}^2 + \sigma^2)^{-\frac{(\nu-m)+2}{2}}
\end{align*}

Putting all of these terms together, we arrive at the following full Jacobian of the SVD, inverse, and PCA-reparameterization transformations,
\begin{align*}
     (dY^{(0)}) &= \prod_{j>i}^k \left((L_{jj}^2 + \sigma^2)^{-1} - (L_{ii}^2 + \sigma^2)^{-1}\right)
     \left(\prod_{j=1}^{k} ((L_{jj}^2 + \sigma^2)^{-1} - \sigma^{-2})^{(m-k)} \right)\\
     &\quad \left((\sigma^{2*\frac{1}{2}})^{-(\nu-m)(m-k)-\binom{m-k}{2}-2}\right)
     \left(\prod_{j = 1}^k (L_{jj}^2 + \sigma^2)^{-\frac{(\nu-m)+2}{2}}\right)\\
     &\quad \left(\prod_{j = 1}^{k} (L_{jj}^2 + \sigma^2)^{-\frac{1}{2}}L_{jj}\right) \frac{1}{2}(\sigma^2)^{-\frac{1}{2}}(dL)(d\sigma^2)\\
     &\quad \bigwedge_{i = m+1}^\nu \bigwedge_{j=1}^m \Omega_{ij} \bigwedge_{i = k+1}^m \bigwedge_{j=i+1}^m (\Omega_{ij} - \Psi_{ij}) \bigwedge_{i = k+1}^m \bigwedge_{j=1}^k \Omega_{ij}\Psi_{ij}\bigwedge_{i = 1}^k \bigwedge_{j=i+1}^k \Omega_{ij}\Psi_{ij}\\
     &= \prod_{j>i}^k \left((L_{jj}^2 + \sigma^2)^{-1} - (L_{ii}^2 + \sigma^2)^{-1}\right)
     \left(\prod_{j=1}^{k} ((L_{jj}^2 + \sigma^2)^{-1} - \sigma^{-2})^{(m-k)} \right)\\
     &\quad \left((\sigma^{2*\frac{1}{2}})^{-(\nu-m)(m-k)-\binom{m-k}{2}-3}\right)
     \left(\prod_{j = 1}^k (L_{jj}^2 + \sigma^2)^{-\frac{(\nu-m)+3}{2}}\right)\\
     &\quad \frac{1}{2} \det(L) (dL)(d\sigma^2)\\
     &\quad \bigwedge_{i = m+1}^\nu \bigwedge_{j=1}^m \Omega_{ij} \bigwedge_{i = k+1}^m \bigwedge_{j=i+1}^m (\Omega_{ij} - \Psi_{ij}) \bigwedge_{i = k+1}^m \bigwedge_{j=1}^k \Omega_{ij}\Psi_{ij}\bigwedge_{i = 1}^k \bigwedge_{j=i+1}^k \Omega_{ij}\Psi_{ij}
\end{align*}
\subsection{Derivation of prior distribution}
Following the above sections, we assume the $\nu\times m$ dimensional $Y^{(0)} \sim N(0,\Phi)$ where $\Phi$ is an $m\times m$ spatial covariance matrix and only the first $k+1$ singular values of $Y^{(0)}$ are unique. Define $\Sigma \coloneqq \Phi^{-1}$ as reinterpreting the precision $\Phi^{-1}$ as a covariance parameter $\Sigma$ after the transformation $J(D\to B^{-1})$.  Let $D_0$ be the first $m$ rows of $D$, $B_0$ be the first $m$ rows of $B$, and $U_0$ be the first $k$ columns of $U$. Conditioned on these assumptions, we proceed with the three transformations (SVD, inverse of $D_0$, PCA reparameterization) and plug in the Jacobians derived above at each of the following steps,
\begin{align*}
    p(Y^{(0)}) (dY^{(0)}) &\propto \exp{\left(-\frac{1}{2}\tr(Y^{(0)'}Y^{(0)} \Phi^{-1})\right)} (dY^{(0)})\\
    &\propto \exp{\left(-\frac{1}{2}\tr(U D_0^2 U' \Phi^{-1})\right)}\\
    &\quad \prod_{j>i}^k \left(D_{jj}^2-D_{ii}^2\right)\prod_{j=1}^{k} (D_{jj}^2 - D_{k+1,k+1}^2)^{(m-k)} D_{k+1,k+1}^{(\nu-m)(m-k)+\binom{m-k}{2}} \prod_{j = 1}^k D_{jj}^{\nu-m}(dD)\\ 
    &\propto \exp{\left(-\frac{1}{2}\tr(U B_0^{-2} U' \Sigma)\right)}\\
    &\quad \prod_{j>i}^k \left(B_{jj}^{-2}-B_{ii}^{-2}\right)\prod_{j=1}^{k} (B_{jj}^{-2} - B_{k+1,k+1}^{-2})^{(m-k)} B_{k+1,k+1}^{-(\nu-m)(m-k)-\binom{m-k}{2}-2}\\
    &\quad \prod_{j = 1}^k B_{jj}^{-(\nu-m)-2} (dB)\\
    &\propto \exp{\left(-\frac{1}{2}\tr(U \begin{bmatrix}
        (L^2 + \sigma^2 I_k)^{\frac{1}{2}} & 0_{k,m-k}\\ 0_{m-k,k} & (\sigma^2)^{\frac{1}{2}}I_{m-k}
    \end{bmatrix}^{-2} U' \Sigma)\right)}\\
    &\quad \prod_{j>i}^k \left((L_{jj}^2 + \sigma^2)^{-1} - (L_{ii}^2 + \sigma^2)^{-1}\right)
     \left(\prod_{j=1}^{k} ((L_{jj}^2 + \sigma^2)^{-1} - \sigma^{-2})^{(m-k)} \right)\\
     &\quad \left((\sigma^{2*\frac{1}{2}})^{-(\nu-m)(m-k)-\binom{m-k}{2}-3}\right)
     \left(\prod_{j = 1}^k (L_{jj}^2 + \sigma^2)^{-\frac{(\nu-m)+3}{2}}\right)\\
     &\quad \frac{1}{2} \det(L) (dL)(d\sigma^2)\\
\end{align*}
Finally, redefining $U \coloneqq U_{(1:k)}$ as the first $k$ columns of the full orthogonal matrix, we can rewrite the prior in the form used for inference,
\begin{align*}
    p(U,L, \sigma^2|\Sigma) &\propto \exp{\left(-\frac{1}{2}\tr(U \left((L^2 + \sigma^2 I_k)^{-1} - \sigma^{-2}I_k\right) U' \Sigma + \frac{1}{\sigma^2}\Sigma)\right)}\\
    &\quad \prod_{j>i}^k \left((L_{jj}^2 + \sigma^2)^{-1} - (L_{ii}^2 + \sigma^2)^{-1}\right)
     \left(\prod_{j=1}^{k} ((L_{jj}^2 + \sigma^2)^{-1} - \sigma^{-2})^{(m-k)} \right)\\
     &\quad \left((\sigma^{2*\frac{1}{2}})^{-(\nu-m)(m-k)-\binom{m-k}{2}-3}\right)
     \left(\prod_{j = 1}^k (L_{jj}^2 + \sigma^2)^{-\frac{(\nu-m)+3}{2}}\right)\\
     &\quad \frac{1}{2} \det(L) (dL)(d\sigma^2)\\
\end{align*}

\subsection{Proof of Theorem \ref{thm:MAP_U}}
This section contains the proof of Theorem \ref{thm:MAP_U}.
\subsubsection{MAP inference of orthogonal matrix U}
Let $S = Y'Y/n$, where $Y$ has dimension $n\times m$. The conditional (and marginal over $\{z_i\}$) posterior distribution of $U$ is proportional to,
\begin{align*}
p(U|Y, L,\sigma^2,\Sigma) &\propto |UL^2U' + \sigma^2 I_m|^{-\frac{n}{2}} \etr(-\frac{1}{2} nS(UL^2U' + \sigma^2 I_m)^{-1})\\ 
&\quad\quad\etr(-\frac{1}{2}\Sigma U\left((L^2+\sigma^2I_k)^{-1}-\sigma^{-2}I_k)\right)U'
\end{align*}
By the Woodbury Identity on the inverse covariance from the likelihood,
\[
(UL^2U' + \sigma^2 I_m)^{-1} = \sigma^{-2}I_m - \sigma^{-4}U(L^{-2}+ \sigma^{-2}I_k)^{-1}U'
\]
And another application on the inner inverse,
\[
(L^{-2}+ \sigma^{-2}I_k)^{-1} = \sigma^2 I_k - \sigma^4 (L^2 + \sigma^2 I_k)^{-1}
\]
Putting them together,
\[
(UL^2U' + \sigma^2 I_m)^{-1} = \sigma^{-2}I_m + U\left((L^2 + \sigma^2 I_k)^{-1}- \sigma^{-2} I_k\right)U'
\]
Which leads to the proportionality of the posterior of interest,
\begin{align*}
p(U|Y, L,\sigma^2,\Sigma) &\propto |UL^2U' + \sigma^2 I_m|^{-\frac{n}{2}} \etr(-\frac{1}{2}(nS+\Sigma) U\left((L^2+\sigma^2I_k)^{-1}-\sigma^{-2}I_k)\right)U'
\end{align*}
To get rid of the determinant's dependence on $U$, let $\Tilde{U} = \begin{bmatrix} U&U^{\perp} \end{bmatrix}$ for some $U^{\perp}$ to complete the basis as a function of $U$,
\begin{align*}
UL^2U' + \sigma^2 I_m &= \Tilde{U}\begin{bmatrix}
    L^2 & 0_{k,m-k}\\ 0_{m-k,k} & 0_{m-k,m-k}
\end{bmatrix}\Tilde{U}' + \sigma^2 \Tilde{U}\Tilde{U}'\\
&= \Tilde{U}\begin{bmatrix}
    L^2+\sigma^2 I_k & 0_{k,m-k}\\ 0_{m-k,k} & \sigma^2 I_{m-k}
\end{bmatrix}\Tilde{U}'
\end{align*}
Thus, the posterior is proportional to,
\begin{align*}
p(U|Y, L,\sigma^2,\Sigma) &\propto \etr(-\frac{1}{2}(nS+\Sigma) U\left((L^2+\sigma^2I_k)^{-1}-\sigma^{-2}I_k\right)U')
\end{align*}
The MAP value of this distribution is evaluated at,
\[
U_{map} = \underset{U}{\mathrm{argmax}}\ \tr(U'(nS+\Sigma) U\left(\sigma^{-2}I_k - (L^2+\sigma^2I_k)^{-1}\right))
\]
By \cite{absil2008optimization}, section 4.8.2, the critical points of this function are eigenvectors of $(nS + \Sigma)$, and thus it remains to be proved which eigenvectors are the MAP. Note, since the diagonal entries of $L$ are decreasing, the diagonal entries of $(\sigma^{-2}I_k - (L^2+\sigma^2I_k)^{-1})$ are decreasing. Furthermore, since $\Psi \coloneqq U'(nS+\Sigma) U$ is a diagonal matrix of eigenvalues of $(nS+\Sigma)$, maximizing the trace amounts to choosing the largest eigenvalues in decreasing order by the rearrangement inequality.

\section{Appendix: Laminae-PCA}
\label{sec:laminae_pca}
In this appendix, we briefly introduce the Laminae-PCA model used in the DLPFC case study.  The defining feature of Laminae-PCA is that the spatial loadings $U$ are fixed to the normalized layer-indicator matrix constructed from the manual cortical layer and white matter annotations provided by \citet{maynard2021transcriptome}. This model is useful as a supervised anatomical reference for characterizing gene expression coefficients under this structural assumption. 

Let $\Pi \in \{0,1\}^{(m\times k)}$ be the laminar partition, such that the sum of the columns of $\Pi$ is an $m$-length vector of all $1$'s. Then define $U$ as the normalization of $\Pi$, such that each column of $\Pi$ is divided by its norm.  Then $U$ is orthonormal, i.e. $U'U = I_k$ since $\Pi$ was a partition. Using this construction of the loadings, a probabilistic PCA model follows as usual,
\[
\bm{y}_i | U,L,\sigma^2,\bm{z}_i \sim N(UL\bm{z}_i, \sigma^2 I_m)
\]
In this setting, the E-step for $\bm{z}_i$ does not change, and maximizing the $Q$ function within the EM framework with respect to $L$ and $\sigma^2$ can be carried out using autograd.
\section{Appendix: Percent of Variance Explained}
\label{sec:appendix_pve}
We use percent of variance explained as a metric in our case studies to understand the performance of PCA, SOFM, and Laminae-PCA with respect to the empirical covariance matrix as well as the prior covariance matrix. Let $R$ be the covariance matrix of interest and let $V$ be the loadings of a specific model. Then, the percent of variance explained is calculated via,
\[
pve = \frac{\tr(V'RV)}{\tr(R)}
\]
Note that in our analysis, $R$ can be the empirical covariance matrix, the spatial prior, or the Laminae-PCA model covariance, and the orthonormal $V$ matrix can be from the PCA, SOFM, or Laminae-PCA models. By interchanging these, we can learn how close these models are to each other. For example, no model can perform better than PCA on the empirical covariance matrix $S$ in terms of $pve$ \citep{jolliffe2016principal}.
\section{Appendix: Additional simulation study results}
\label{sec:app_synth}
This appendix contains additional simulation studies. 
\begin{figure}[htbp]
    \centering
    \includegraphics[width=\textwidth]{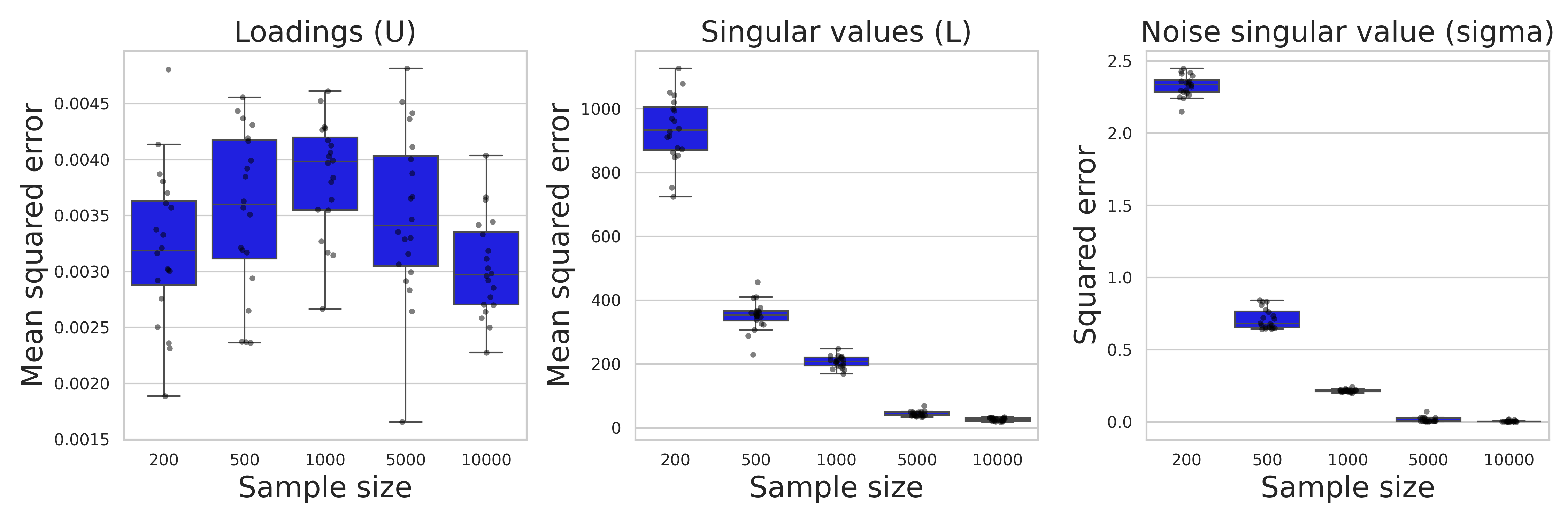}
    \caption{Synthetic study of nonstationary model performance on data generated with a spatial covariance from the circular model, with $\sigma = 10.0$, a maximum lag of 20 and size of spatial domain equal to 400.}
\end{figure}

% \begin{figure}[htbp]
%     \centering
%     \includegraphics[width=\textwidth]{synth_study_figs/sofm_latlonrot_plots__sigma10.0_maxlag20_growingTrue_priorcov0.png}
%     \caption{Synthetic study of nonstationary model performance on data generated with a spatial covariance from the circular model, with $\sigma = 10.0$, a maximum lag of 20 and size of spatial domain equal to the sample size.}
% \end{figure}

\begin{figure}[htbp]
    \centering
    \includegraphics[width=\textwidth]{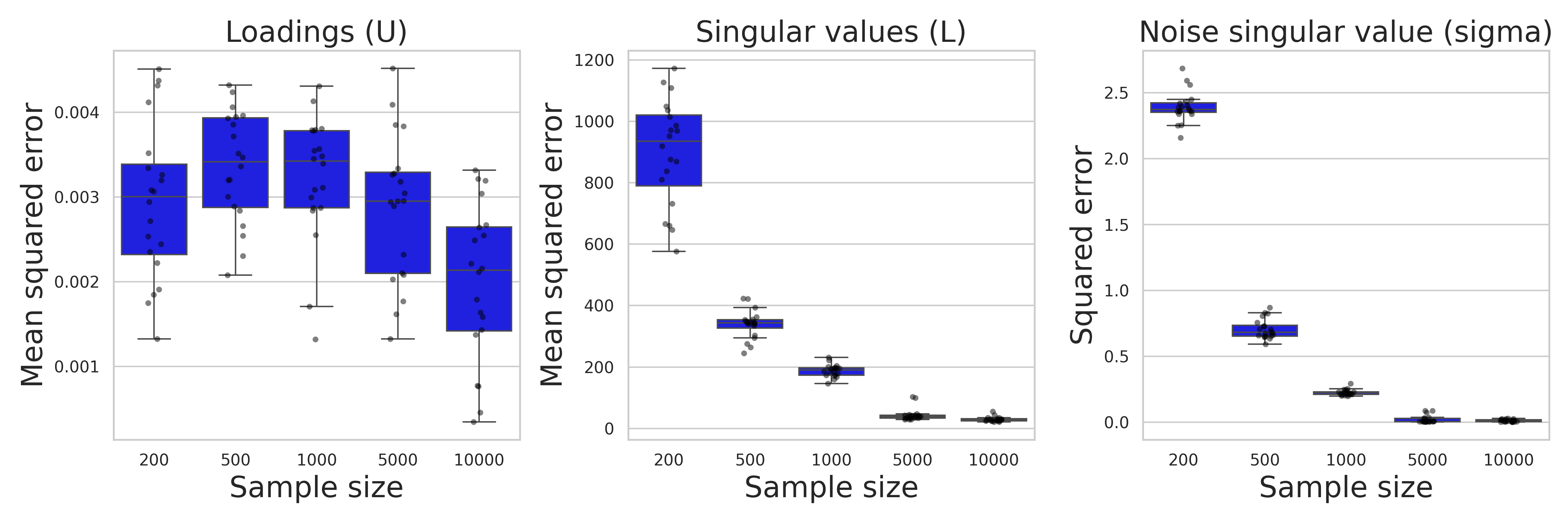}
    \caption{Synthetic study of nonstationary model performance on data generated with a spatial covariance from the stationary model, with $\sigma = 10.0$, a maximum lag of 20 and size of spatial domain equal to 400.}
\end{figure}

% \begin{figure}[htbp]
%     \centering
%     \includegraphics[width=\textwidth]{synth_study_figs/sofm_latlonrot_plots__sigma10.0_maxlag20_growingTrue_priorcov1.png}
%     \caption{Synthetic study of nonstationary model performance on data generated with a spatial covariance from the stationary model, with $\sigma = 10.0$, a maximum lag of 20 and size of spatial domain equal to the sample size.}
% \end{figure}

\begin{figure}[htbp]
    \centering
    \includegraphics[width=\textwidth]{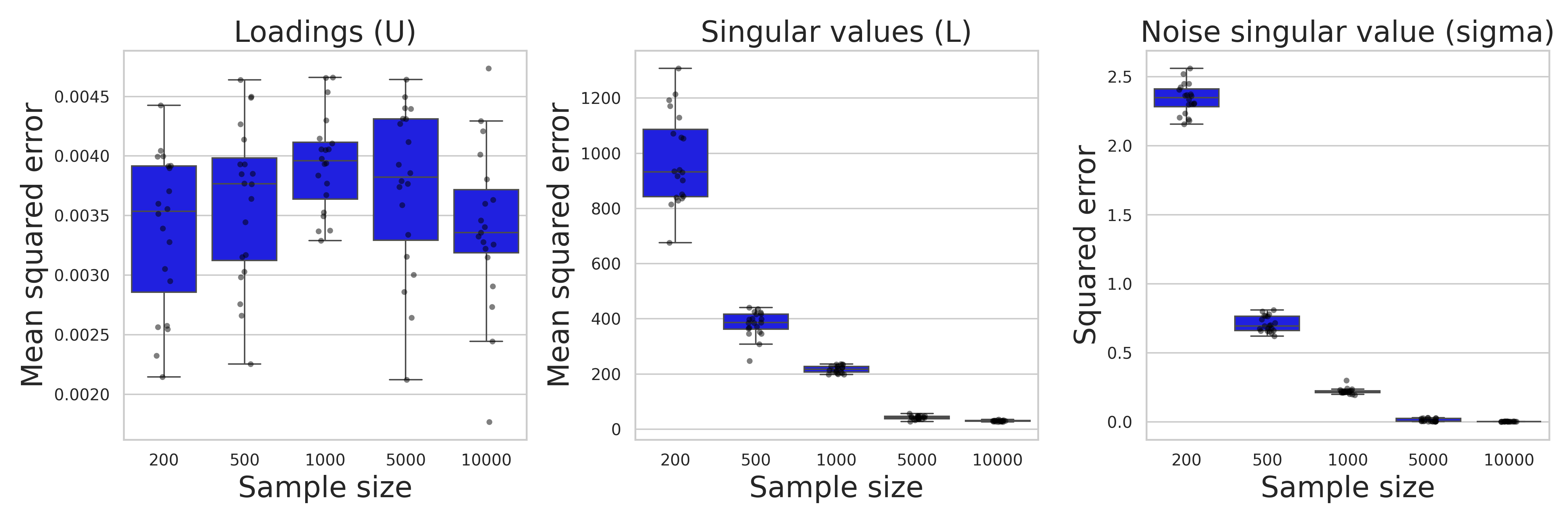}
    \caption{Synthetic study of nonstationary model performance on data generated with a spatial covariance from the left-to-right model, with $\sigma = 10.0$, a maximum lag of 20 and size of spatial domain equal to 400.}
\end{figure}

% \begin{figure}[htbp]
%     \centering
%     \includegraphics[width=\textwidth]{synth_study_figs/sofm_latlonrot_plots__sigma10.0_maxlag20_growingTrue_priorcov2.png}
%     \caption{Synthetic study of nonstationary model performance on data generated with a spatial covariance from the left-to-right model, with $\sigma = 10.0$, a maximum lag of 20 and size of spatial domain equal to the sample size.}
% \end{figure}

\begin{figure}[htbp]
    \centering
    \includegraphics[width=\textwidth]{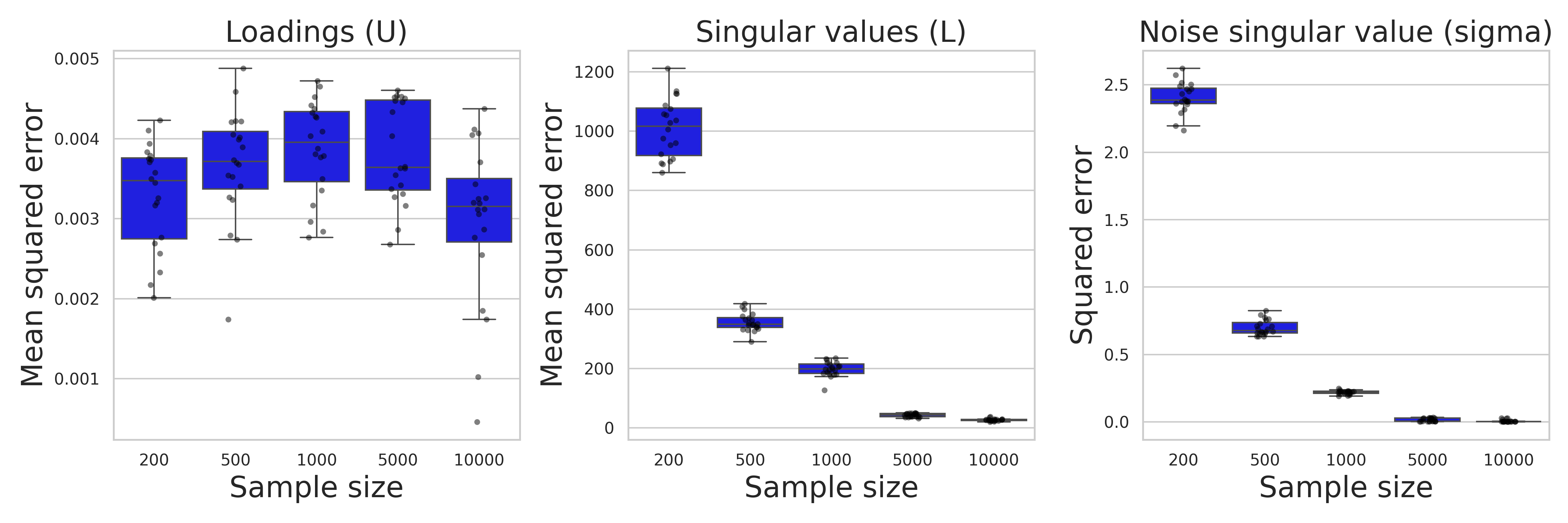}
    \caption{Synthetic study of nonstationary model performance on data generated with a spatial covariance from the circular model, with $\sigma = 10.0$, a maximum lag of 2 and size of spatial domain equal to 400.}
\end{figure}

% \begin{figure}[htbp]
%     \centering
%     \includegraphics[width=\textwidth]{synth_study_figs/sofm_latlonrot_plots__sigma10.0_maxlag2_growingTrue_priorcov0.png}
%     \caption{Synthetic study of nonstationary model performance on data generated with a spatial covariance from the circular model, with $\sigma = 10.0$, a maximum lag of 2 and size of spatial domain equal to the sample size.}
% \end{figure}

\begin{figure}[htbp]
    \centering
    \includegraphics[width=\textwidth]{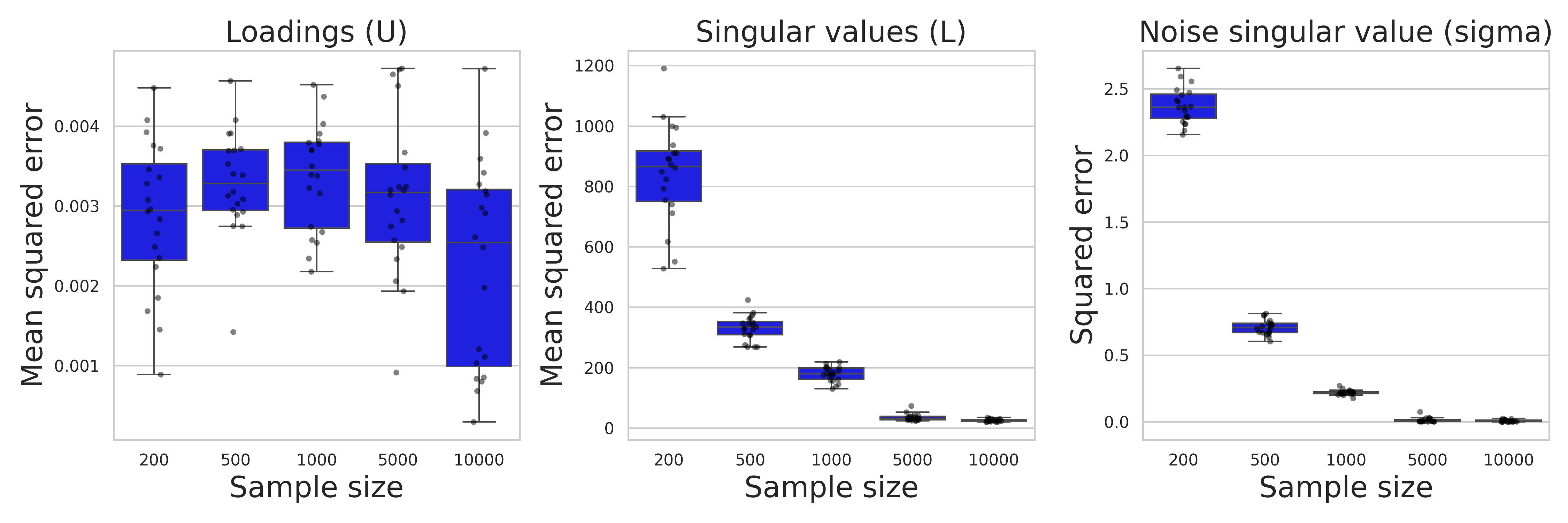}
    \caption{Synthetic study of nonstationary model performance on data generated with a spatial covariance from the stationary model, with $\sigma = 10.0$, a maximum lag of 2 and size of spatial domain equal to 400.}
\end{figure}

% \begin{figure}[htbp]
%     \centering
%     \includegraphics[width=\textwidth]{synth_study_figs/sofm_latlonrot_plots__sigma10.0_maxlag2_growingTrue_priorcov1.png}
%     \caption{Synthetic study of nonstationary model performance on data generated with a spatial covariance from the stationary model, with $\sigma = 10.0$, a maximum lag of 2 and size of spatial domain equal to the sample size.}
% \end{figure}

\begin{figure}[htbp]
    \centering
    \includegraphics[width=\textwidth]{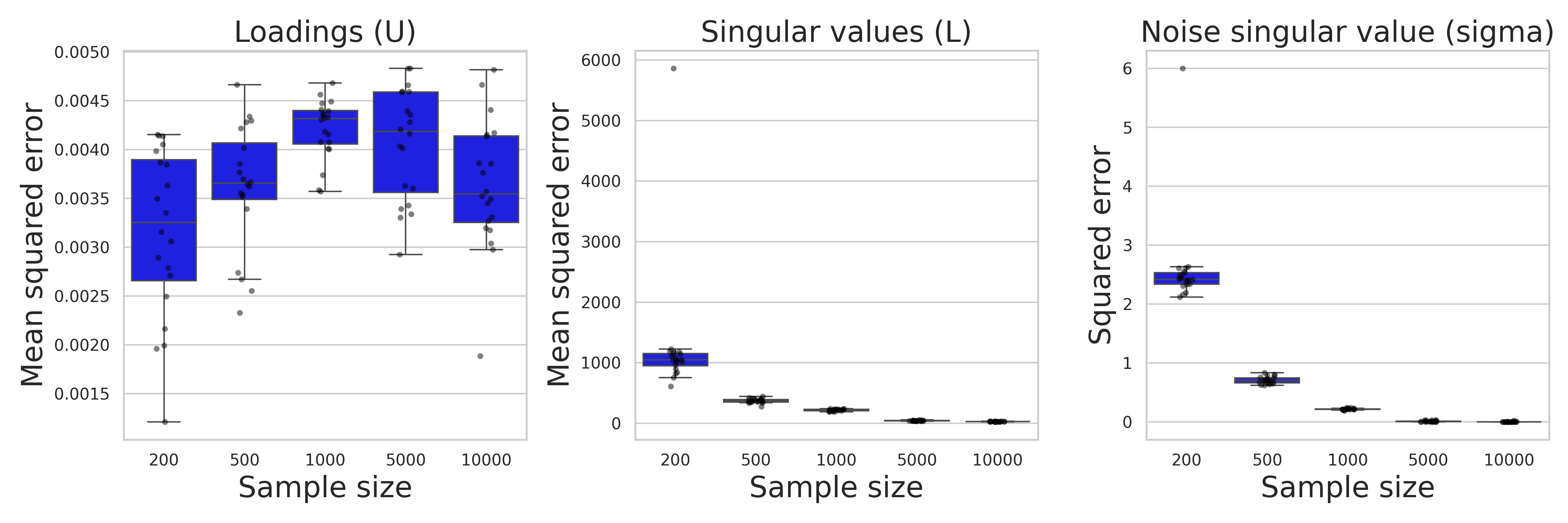}
    \caption{Synthetic study of nonstationary model performance on data generated with a spatial covariance from the left-to-right model, with $\sigma = 10.0$, a maximum lag of 2 and size of spatial domain equal to 400.}
\end{figure}

% \begin{figure}[htbp]
%     \centering
%     \includegraphics[width=\textwidth]{synth_study_figs/sofm_latlonrot_plots__sigma10.0_maxlag2_growingTrue_priorcov2.png}
%     \caption{Synthetic study of nonstationary model performance on data generated with a spatial covariance from the left-to-right model, with $\sigma = 10.0$, a maximum lag of 2 and size of spatial domain equal to the sample size.}
% \end{figure}

\begin{figure}[htbp]
    \centering
    \includegraphics[width=\textwidth]{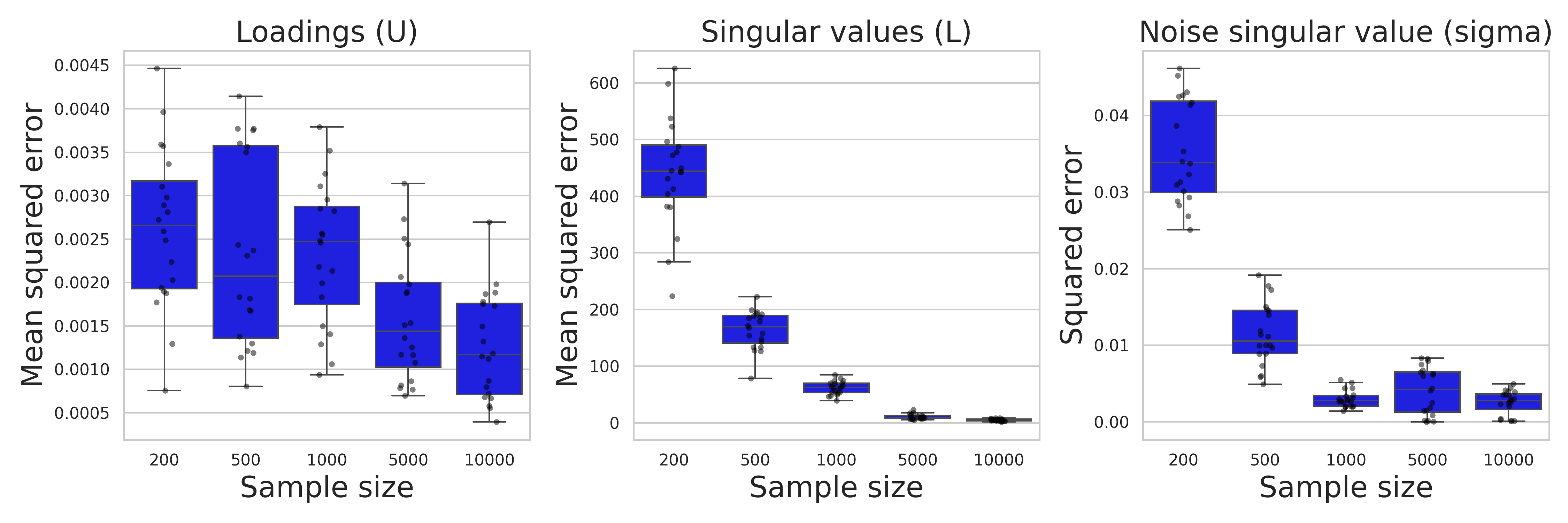}
    \caption{Synthetic study of nonstationary model performance on data generated with a spatial covariance from the circular model, with $\sigma = 5.0$, a maximum lag of 20 and size of spatial domain equal to 400.}
\end{figure}

% \begin{figure}[htbp]
%     \centering
%     \includegraphics[width=\textwidth]{synth_study_figs/sofm_latlonrot_plots__sigma5.0_maxlag20_growingTrue_priorcov0.png}
%     \caption{Synthetic study of nonstationary model performance on data generated with a spatial covariance from the circular model, with $\sigma = 5.0$, a maximum lag of 20 and size of spatial domain equal to the sample size.}
% \end{figure}

\begin{figure}[htbp]
    \centering
    \includegraphics[width=\textwidth]{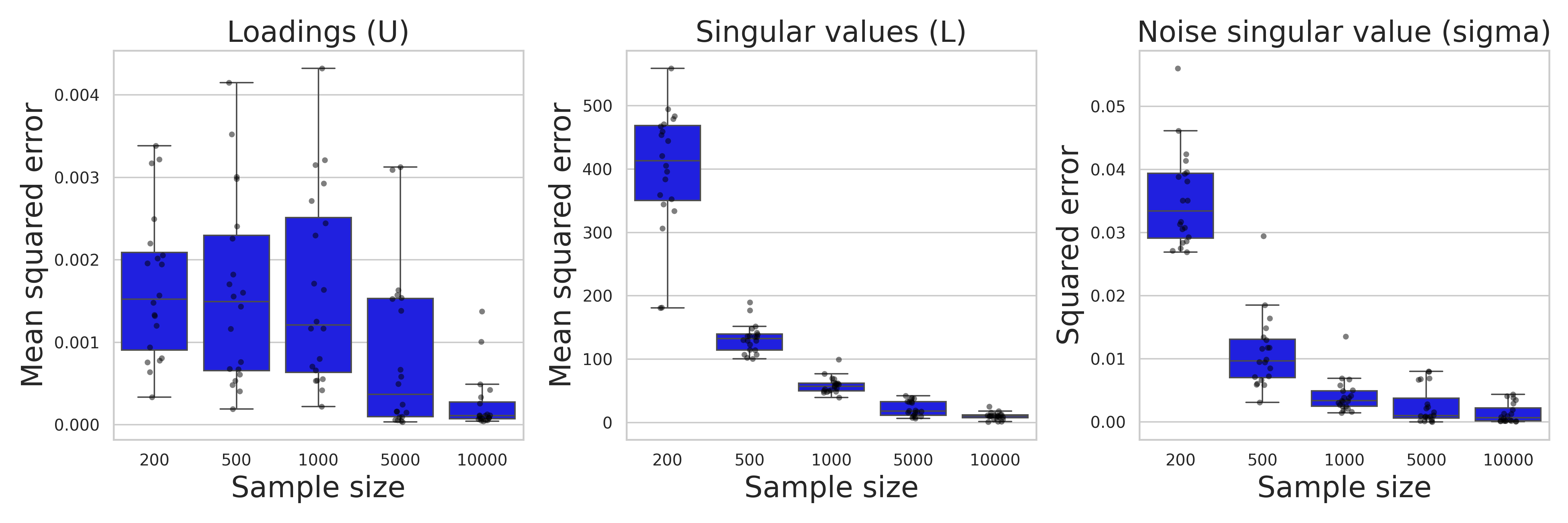}
    \caption{Synthetic study of nonstationary model performance on data generated with a spatial covariance from the stationary model, with $\sigma = 5.0$, a maximum lag of 20 and size of spatial domain equal to 400.}
\end{figure}

% \begin{figure}[htbp]
%     \centering
%     \includegraphics[width=\textwidth]{synth_study_figs/sofm_latlonrot_plots__sigma5.0_maxlag20_growingTrue_priorcov1.png}
%     \caption{Synthetic study of nonstationary model performance on data generated with a spatial covariance from the stationary model, with $\sigma = 5.0$, a maximum lag of 20 and size of spatial domain equal to the sample size.}
% \end{figure}

\begin{figure}[htbp]
    \centering
    \includegraphics[width=\textwidth]{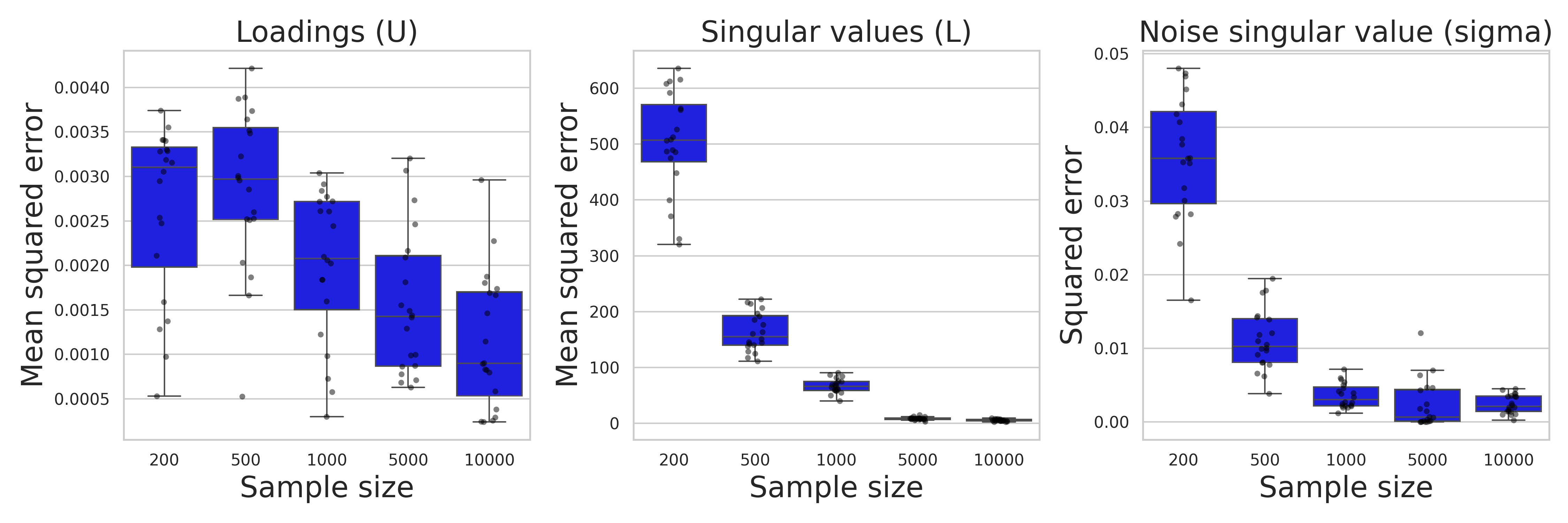}
    \caption{Synthetic study of nonstationary model performance on data generated with a spatial covariance from the left-to-right model, with $\sigma = 5.0$, a maximum lag of 20 and size of spatial domain equal to 400.}
\end{figure}

% \begin{figure}[htbp]
%     \centering
%     \includegraphics[width=\textwidth]{synth_study_figs/sofm_latlonrot_plots__sigma5.0_maxlag20_growingTrue_priorcov2.png}
%     \caption{Synthetic study of nonstationary model performance on data generated with a spatial covariance from the left-to-right model, with $\sigma = 5.0$, a maximum lag of 20 and size of spatial domain equal to the sample size.}
% \end{figure}

\begin{figure}[htbp]
    \centering
    \includegraphics[width=\textwidth]{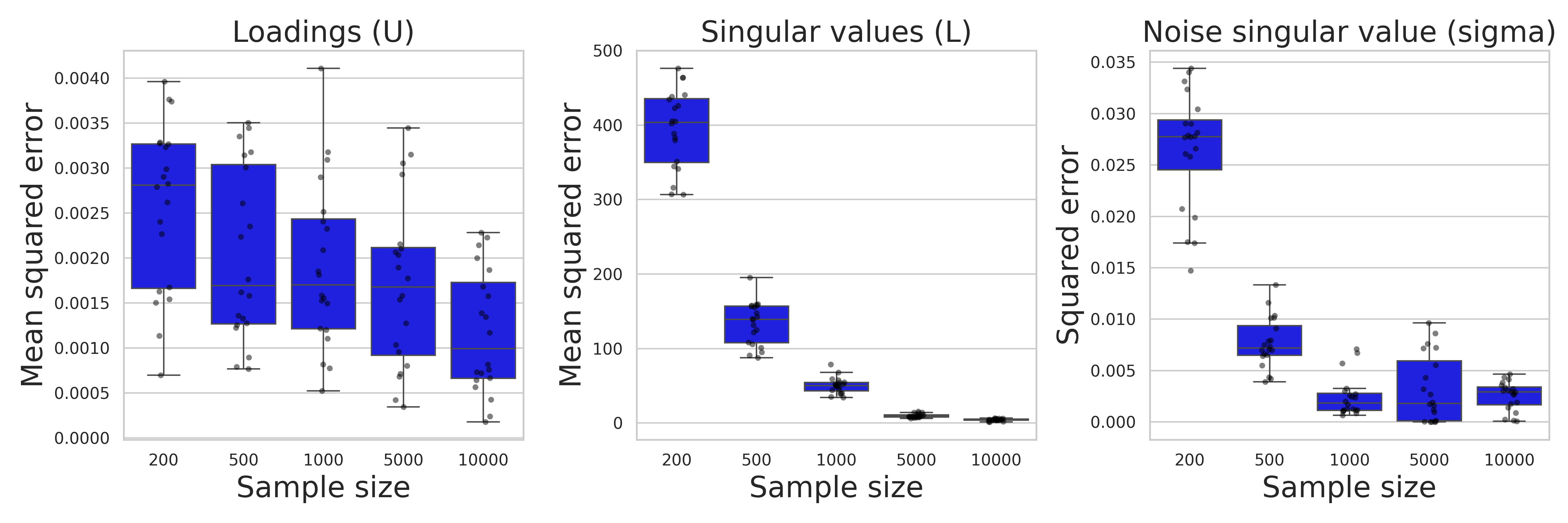}
    \caption{Synthetic study of nonstationary model performance on data generated with a spatial covariance from the circular model, with $\sigma = 5.0$, a maximum lag of 2 and size of spatial domain equal to 400.}
\end{figure}

% \begin{figure}[htbp]
%     \centering
%     \includegraphics[width=\textwidth]{synth_study_figs/sofm_latlonrot_plots__sigma5.0_maxlag2_growingTrue_priorcov0.png}
%     \caption{Synthetic study of nonstationary model performance on data generated with a spatial covariance from the circular model, with $\sigma = 5.0$, a maximum lag of 2 and size of spatial domain equal to the sample size.}
% \end{figure}

\begin{figure}[htbp]
    \centering
    \includegraphics[width=\textwidth]{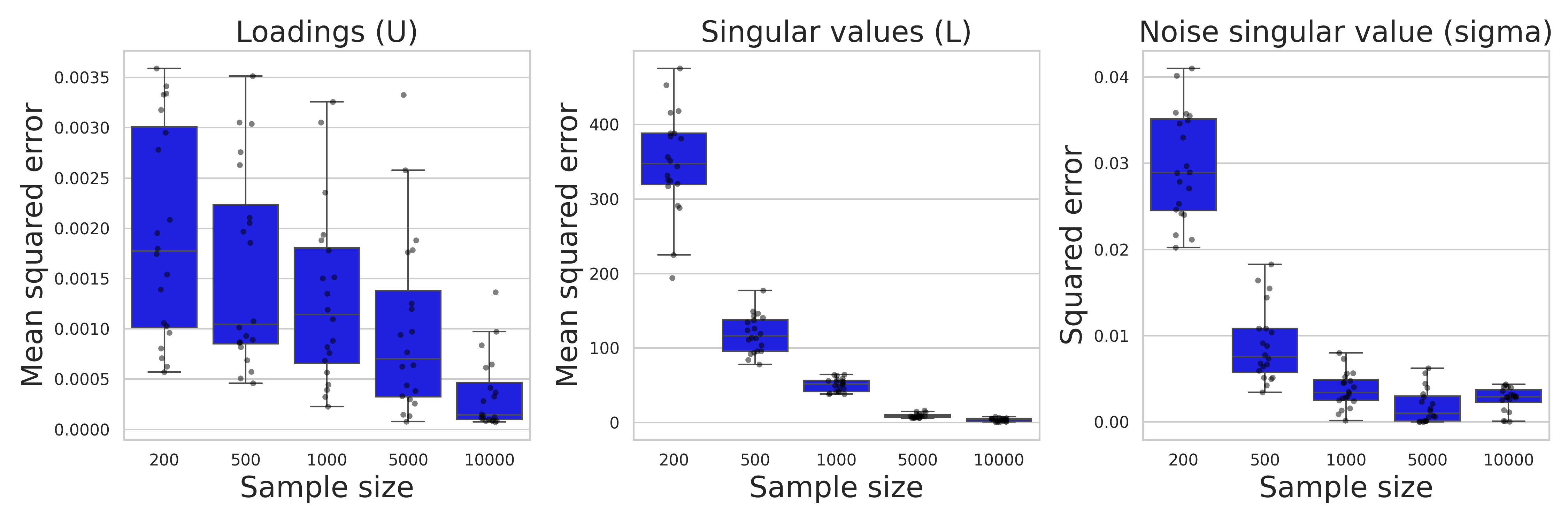}
    \caption{Synthetic study of nonstationary model performance on data generated with a spatial covariance from the stationary model, with $\sigma = 5.0$, a maximum lag of 2 and size of spatial domain equal to 400.}
\end{figure}

% \begin{figure}[htbp]
%     \centering
%     \includegraphics[width=\textwidth]{synth_study_figs/sofm_latlonrot_plots__sigma5.0_maxlag2_growingTrue_priorcov1.png}
%     \caption{Synthetic study of nonstationary model performance on data generated with a spatial covariance from the stationary model, with $\sigma = 5.0$, a maximum lag of 2 and size of spatial domain equal to the sample size.}
% \end{figure}

\begin{figure}[htbp]
    \centering
    \includegraphics[width=\textwidth]{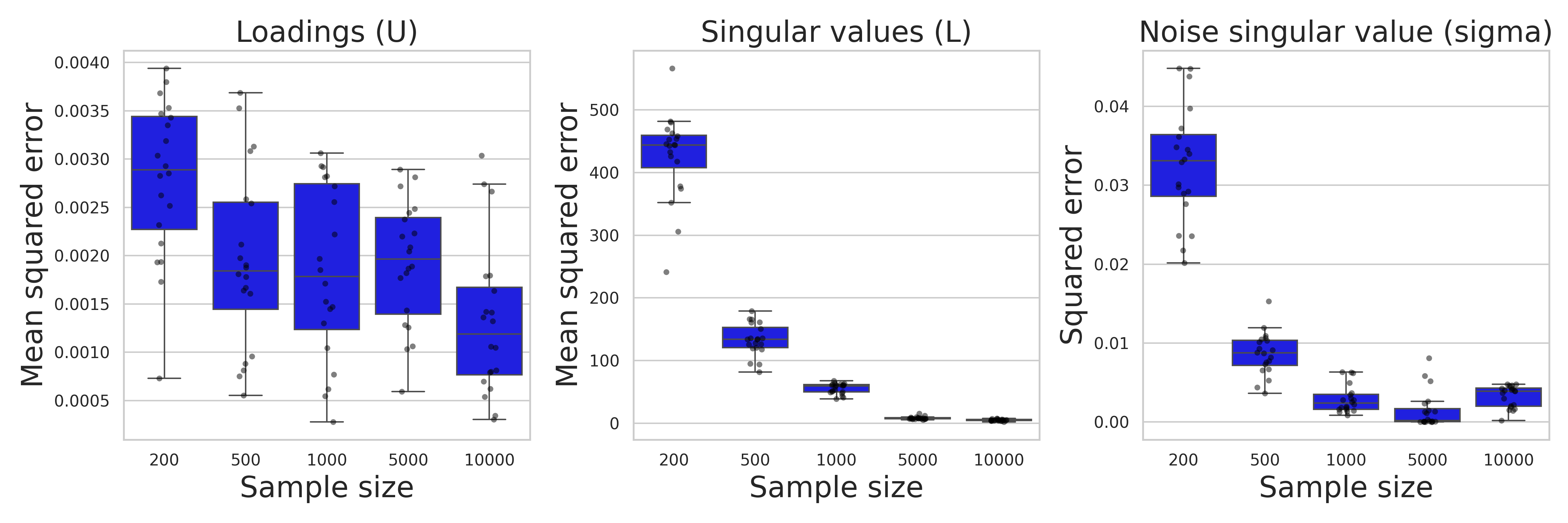}
    \caption{Synthetic study of nonstationary model performance on data generated with a spatial covariance from the left-to-right model, with $\sigma = 5.0$, a maximum lag of 2 and size of spatial domain equal to 400.}
\end{figure}

% \begin{figure}[htbp]
%     \centering
%     \includegraphics[width=\textwidth]{synth_study_figs/sofm_latlonrot_plots__sigma5.0_maxlag2_growingTrue_priorcov2.png}
%     \caption{Synthetic study of nonstationary model performance on data generated with a spatial covariance from the left-to-right model, with $\sigma = 5.0$, a maximum lag of 2 and size of spatial domain equal to the sample size.}
% \end{figure}
\end{document}